\documentclass{aa}  

\usepackage{graphicx}
\usepackage[utf8]{inputenc}
\usepackage{txfonts}
\usepackage{geometry, amsmath, float, amssymb, amsfonts, graphicx, enumitem, wrapfig, url, framed, placeins, fancyhdr, wrapfig, color, textcomp}

\begin{document} 
   \title{Solar prominence diagnostics from non-LTE modelling of \ion{Mg}{ii} h\&k line profiles
   }
   \author{A. W. Peat \inst{1},
          N. Labrosse \inst{1},
          B. Schmieder \inst{1,2,3}, \and
          K. Barczynski \inst{2,4,5} 
          } 

   \institute{SUPA School of Physics and Astronomy, The University of Glasgow,
              Glasgow, G12 8QQ, UK\\
              \email{a.peat.1@research.gla.ac.uk} \and LESIA, Observatoire de Paris, Universit\'e PSL , CNRS, Sorbonne Universit\'e, Universit\'e Paris-Diderot, 5 place Jules Janssen, 92190 Meudon, France \and KU Leuven, Belgium \and PMOD/WRC, Dorfstrasse 33, CH-7260 Davos Dorf, Switzerland \and ETH-Zurich, H\"onggerberg campus, HIT building, Z\"urich, Switzerland}

   \date{}

 
  \abstract
   {}
   {We investigate a new method to for obtaining the plasma parameters of solar prominences observed in the \ion{Mg}{ii} h\&k spectral lines by comparing line profiles from the IRIS satellite to a bank of profiles computed with a one-dimensional non-local thermodynamic equilibrium (non-LTE) radiative transfer code. }
   {Using a grid of 1007 one-dimensional non-LTE radiative transfer models, some including a prominence-corona transition region (PCTR), we carry out this new method to match computed spectra to observed line profiles while accounting for line core shifts not present in the models. The prominence observations were carried out by the IRIS satellite on 19 April 2018.}
   {The prominence is very dynamic with many flows, including a large arm extending from the main body seen near the end of the observation. This flow is found to be redshifted, as is the prominence overall. The models are able to recover satisfactory matches in areas of the prominence where single line profiles are observed. We recover: mean temperatures of 6000 to 50~000K; mean pressures of 0.01 to 0.5 dyne cm$^{-2}$; column masses of 3.7$\times10^{-8}$ to 5$\times10^{-4}$ g cm$^{-2}$; a mean electron density of 7.3$\times10^{8}$ to 1.8$\times10^{11}$ cm$^{-3}$; and an ionisation degree \({n_\text{HII}}/{n_\text{HI}}=0.03 - 4500\). The highest values for the ionisation degree are found in areas where the line of sight crosses mostly plasma from the PCTR, correlating with high mean temperatures and correspondingly no H\(\alpha\) emission.}
   {This new method naturally returns information on how closely the observed and computed profiles match, allowing the user to identify areas where no satisfactory match between models and observations can be obtained. The inclusion of the PCTR was found to be important when fitting models to data as regions where satisfactory fits were found were more likely to contain a model encompassing a PCTR. The line core shift can also be recovered from this new method, and it shows a good qualitative match with that of the line core shift found by the quantile method. This demonstrates the effectiveness of the approach to line core shifts in the new method.}

   \authorrunning{A. W. Peat et al.}
   \maketitle
%

\section{Introduction}
Solar prominences, filaments when viewed on disc, are structures in the solar corona comprising of relatively cool and dense plasma. They form in filament channels, which manifest above polarity inversion lines seen in photospheric magnetograms, suspended in dips in the magnetic field \citep{promreviewi, promreviewii, newprom}. The high spatial and temporal resolution of space-based observations opens a new era of study into the dynamics, evolution, and structure of solar prominences. These properties are key to our understanding of prominence physics.

Since the launch of the Solar Dynamics Observatory \citep[SDO;][]{sdo}, studies of extreme ultraviolet (EUV) prominence observations with the Atmospheric Imaging Assembly \citep[AIA;][]{aia} have revealed new information about prominence barbs and the prominence-corona transition region \citep[PCTR;][]{parenti2012, ruan2018}. The Interface Region Imaging Spectrograph \citep[IRIS;][]{iris} allows for the acquisition of spectroscopic observations of prominences in the near ultraviolet (NUV) and far ultraviolet (FUV). Many past studies have focused on the information one is able to ascertain from statistics of the line profiles \citep{heinzel2014, ruan2018, kucera2018, levenspromlegs, liu2015}.

Recent studies have focused on attempts to invert solar prominence atmospheres to obtain diagnostic information pertaining to that of the atmospheric conditions from IRIS observations of \ion{Mg}{ii} \citep{jejcic2018, zhangcmepromiris, levensmgii}. Such inversions have been done by generating grids of models and matching them with observational data through different statistical methods. These comparisons between observed and computed spectra are based on integrated line intensities and line widths. However, these studies did not provide detailed information on how closely the models match the observations: a `best match' model is always returned, even if the observed and computed line shapes are very different. \cite{heinzel2015} 
fitted a computed line profile to the mean of six line profiles with a focus on the shape of the line profiles rather than the integrated line intensities and line widths. In this paper, a similar type of inversion is done, but here we look at each line profile separately, point-for-point, using a statistical test to assess where the models yield good comparisons with the observations.

The paper is organised as follows. In Section 2 we present observations of a solar prominence obtained on 19 April 2018. In Section 3, we perform a detailed analysis of the characteristics of the \ion{Mg}{ii} h\&k line profiles observed with IRIS. In Section 4, we present the models used in this work and the method developed to compare models and observations, as well as the results that we obtain. In Section 5 we offer our conclusions. 

\section{Observations}
\begin{figure}
    \centering
    \includegraphics[width=\linewidth]{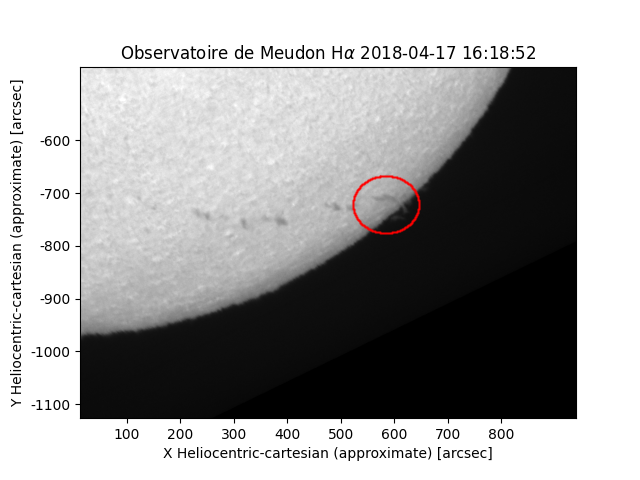}
    \caption{H$\alpha$ observations from the Meudon Spectroheliograph of the solar limb on the days leading up to the coordinated observation. The filament is highlighted with a red circle.}
    \label{meudon}
\end{figure}

\begin{figure}
    \includegraphics[width=0.90\linewidth]{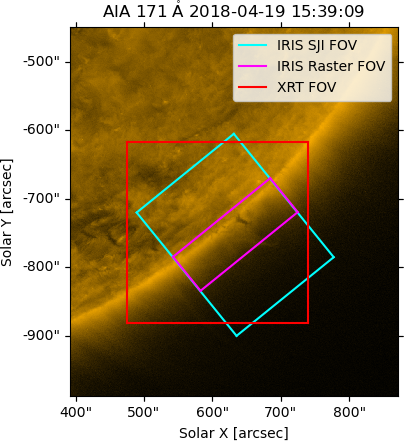}
    \caption{IRIS and XRT (FOV) overlaid on AIA 171~\AA}
    \label{fovs}
\end{figure}

A filament appeared on the south-western solar disc on 17 April 2018 in H$\alpha$ observations from the Meudon Spectroheliograph\footnote{http://bass2000.obspm.fr} (see Figure \ref{meudon}). 
This then later manifested as a prominence off the south-western solar limb on 19 April 2018. The prominence was observed with IRIS, and Hinode \citep{hinodeoverview} as part of a coordinated observation with the Multichannel Subtractive Double Pass (MSDP) spectrograph \citep{mein1991} in the Meudon Solar Tower and other ground-based observatories. The IRIS and Hinode observations start from 14:14 and end at 19:15 UTC. The IRIS observations are comprised of a set of 18 very large coarse 32-step rasters of the \ion{C}{ii} (1331.7~\AA\ to 1358.3~\AA), \ion{Si}{iv} (1388.0~\AA\ to 1406.7~\AA), and \ion{Mg}{ii} (2783.2~\AA\ to 2835~\AA) filters, along with their complimentary slit-jaw imager (SJI) observations centred at 1330~\AA, 1400~\AA, and 2796~\AA, respectively. The rasters had a field of view (FOV) of 63.9\arcsec$\times$182.3\arcsec\ centred on helioprojective coordinates 632.5\arcsec, -753.2\arcsec, with a clockwise satellite rotation angle of 51\textdegree, such that the solar limb was parallel to the y axis of the instruments. The Hinode observations consisted of X-Ray Telescope \citep[XRT;][]{xrt} observations with three filter combinations, Al poly/Open, Open/Gband, and Open/Ti with an FOV of 263.3\arcsec$\times$263.3\arcsec, centred on helioprojective coordinates 607.2\arcsec, -749.7\arcsec. The MSDP observations start at 12:05 UTC and end at 16:35 UTC with a reconstructed FOV of 270\arcsec$\times$370\arcsec \citep[see][]{kb2021}.
AIA whole-disc observations were also available. Figure~\ref{fovs} shows a depiction of the configuration of the AIA, IRIS, and Hinode observations.

\subsection{Hinode/XRT}
\begin{figure}
    \begin{center}
    \includegraphics[width=0.90\linewidth]{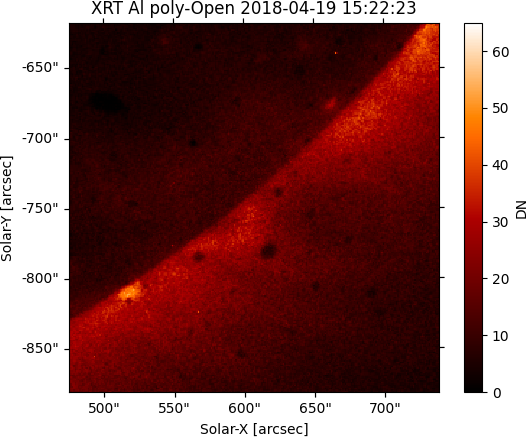}
    \caption{Hinode/XRT Al poly/Open observation of the prominence at 15:22UTC. The coronal cavity in which the prominence sits is clearly seen near (650\arcsec, -750\arcsec).}
    \label{xrt}
    \end{center}
\end{figure}

The images from the Al poly/Open filter of XRT on board Hinode clearly show the coronal cavity in which the prominence sits. However, over the course of the observation, these images do not show any appreciable change or brightenings (see Figure \ref{xrt}).

\subsection{IRIS}
\label{irisobs}
\begin{figure}
    \centering
    \includegraphics[width=\linewidth]{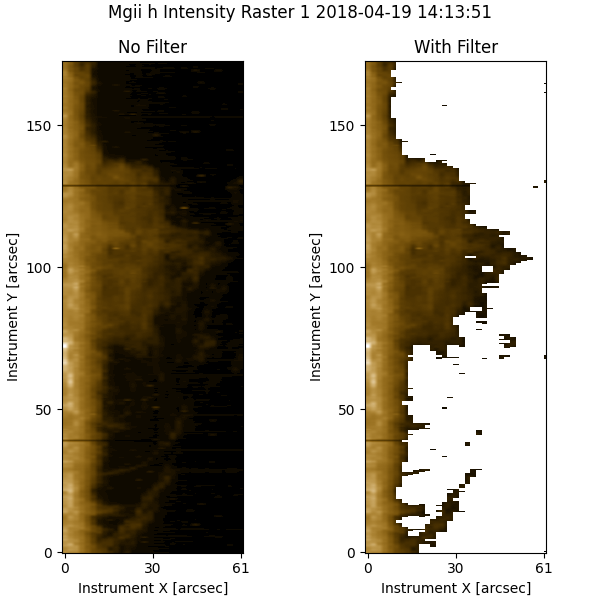}
    \caption{Example of filter. \ion{Mg}{ii} h emission with (right) and without (left) the filter applied to remove noisy and coronal pixel emission.}
    \label{filter}
\end{figure}

The IRIS FITS files were retrieved from the Lockheed Martin Solar and Astrophysics Laboratory (LMSAL) website as level 2 FITS. Radiometric calibration was performed in Python through the method described in \cite{itn26}, with the response function retrieved through \texttt{iris\_get\_response.pro} from SolarSoft (SSW). The data were also deconvolved through Python using a procedure following similar operations to \texttt{iris\_sg\_deconvolve.pro} from SSW. 
Coronal and noisy pixels were removed from the \ion{Mg}{ii} rasters. This was done by calculating a histogram of the full width at half maximum (FWHM) of both \ion{Mg}{ii} h and \ion{Mg}{ii} k in every pixel of every raster. The FWHM was calculated via the quantile method (see section \ref{iris_spec}). These FWHM histograms exhibit a local minimum near the centre of the distributions. Any pixel that exhibits a spectrum with a FWHM above this local minimum is assumed to be a noisy and/or coronal pixel. This can be assumed as random noise will appear to yield a large FWHM via the quantile method. However, some pixels contain spectra with more than one peak, leading to an erroneous measurement of a large FWHM, and some pixels contain spectra that truly have a large FWHM. To combat this, all pixels that fail the FWHM test were then subject to a second test, a simple intensity filter. The limit for this intensity filter was twice the mean of the values of every pixel in every raster in the wavelength range where one would expect \ion{Mg}{ii} h or \ion{Mg}{ii} k emission. Anything above the FWHM limit and below the intensity limit was assumed to be noisy or coronal, and was filtered out. The result of this filtering is presented in Figure \ref{filter}.

In this paper, we focus on the \ion{Mg}{ii} (NUV) filter. The prominence only appears very faintly in the \ion{Si}{iv} and \ion{C}{ii} rasters, making analysis of the FUV bands difficult. We note that reducing the colour-bar limit of the FUV SJI images enough to see the prominence at this low intensity leads to the manifestation of a ghost image of the aperture of the telescope. This is due to parasitic infrared and visible light and only affects off limb observations in the FUV channels; the NUV channel is unaffected \citep{iriscal2018}.

\begin{figure*}
    \centering
    \resizebox{\hsize}{!}
    {\includegraphics[width=0.32\linewidth]{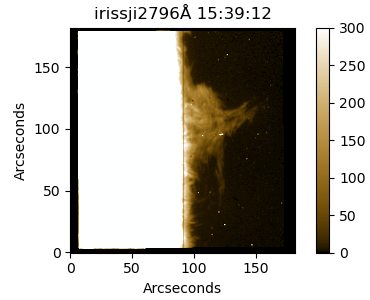}
    \includegraphics[width=0.32\linewidth]{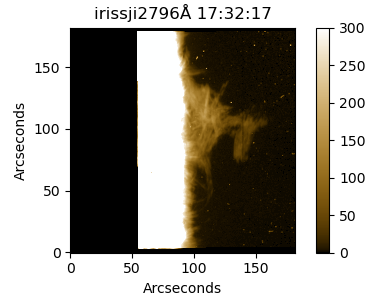}
    \includegraphics[width=0.32\linewidth]{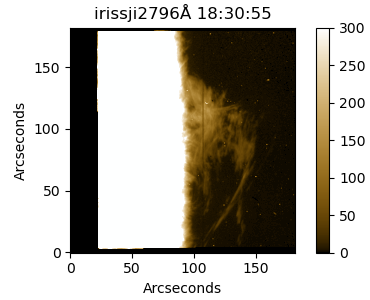}
    }
    \resizebox{\hsize}{!}
    {\includegraphics[width=0.32\linewidth]{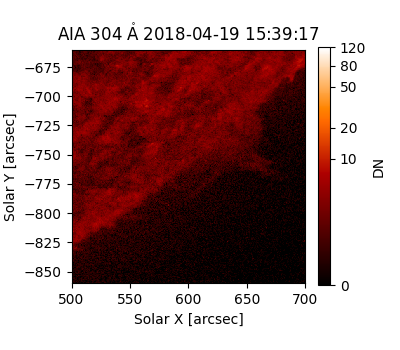}
    \includegraphics[width=0.32\linewidth]{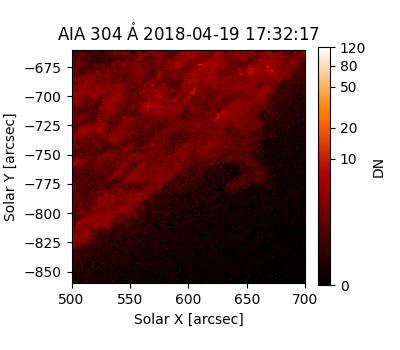}
    \includegraphics[width=0.32\linewidth]{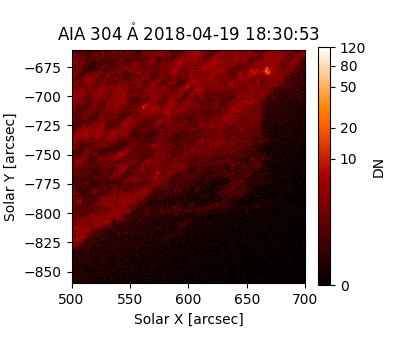}
    }
    \resizebox{\hsize}{!}
    {\includegraphics[width=0.32\linewidth]{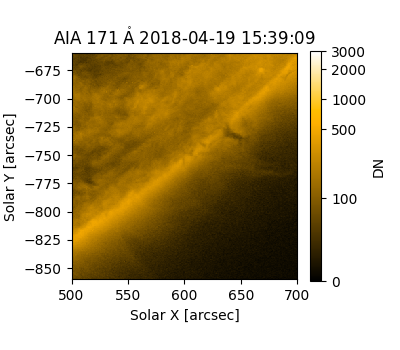}
    \includegraphics[width=0.32\linewidth]{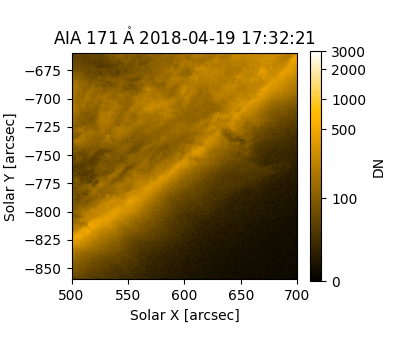}
    \includegraphics[width=0.32\linewidth]{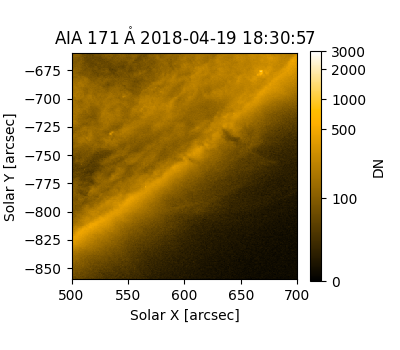}
    }
    \caption{Three main stages of the evolution of the prominence. The top, middle, and bottom rows show, \ion{Mg}{ii}~k IRIS/SJI, 304~\AA\ SDO/AIA, and 171~\AA\ SDO/AIA, respectively. The left, centre, and right columns show small flows from main body, the end of the rise of material, and the large flow of this risen material, respectively.}
    \label{171304}
    
\end{figure*}

In the 2796~\AA\ SJI images, the prominence shows very dynamic behaviour with many flows towards the south-western limb. This culminates in a large flow extending down from a rise of material seen to branch out from the main section of the prominence that persists through the observation. These three stages are shown Figure \ref{171304}.

\subsection{AIA}
The AIA files were retrieved from the Virtual Solar Observatory (VSO) as level 1 FITS. These were then prepared to level 1.5 through the use of \texttt{aia\_prep.pro} from SSW in IDL. The 304~\AA\ filter shows a similar evolution to that seen in the \ion{Mg}{ii} k SJI from IRIS. This is expected as the emission in both channels have some chromospheric origins \citep{iris, aia}. The limb, however, appears higher in 304~\AA\ than in \ion{Mg}{ii}~k; this is likely due to the transition region contribution to the 304~\AA\ filter.

The 171~\AA\ filter clearly shows a small pillar-like barb extending into the region where the prominence is clearly seen in chromospheric filters. This corresponds to the central and densest part of the prominence. One would expect to see strong central reversals in the \ion{Mg}{ii}~h\&k spectra (see Section~\ref{iris_spec}) in this area owing to the high density of material. This is where the prominence appears to be anchored to the surface. In 171~\AA, we also observe a fainter structure enveloping the barb. This can be interpreted as the PCTR since the region of the solar atmosphere imaged by 171~\AA\ includes the upper transition region (TR).

After the end of the IRIS and Hinode observations, the prominence is still seen in the 171~\AA\ and 304~\AA\ filters of AIA. In 304~\AA\ the prominence remains fairly dynamic and appears to fall away towards the limb over the next 24 hours. This can be attributed to the projection effect and the prominence is simply rotating out of view. A similar conclusion is drawn from the 171~\AA\ filter. However, the PCTR and barb of the prominence appear to fade over the next 24 hours as emission from brighter coronal material begins to occult these structures. No filament eruption is observed. On 22 April, the structure reappears in the 171~\AA\ and 304~\AA\ EUVI filters of the Sun Earth Connection Coronal and Heliospheric Investigation \citep[SECCHI;][]{secchi} on board the Solar Terrestrial Relations Observatory (Ahead) \citep[STEREO-A;][]{stereo} and continues to transit across the disc. At this time, the angular separation of STEREO-A from the Earth was approximately 117\textdegree.

\section{IRIS spectra}
\label{iris_spec}
\begin{figure*}[ht]
    \centering
    \includegraphics[width=0.75\linewidth]{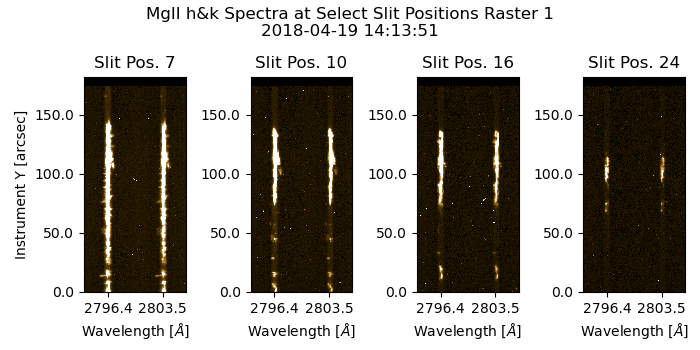}
    \includegraphics[width=0.75\linewidth]{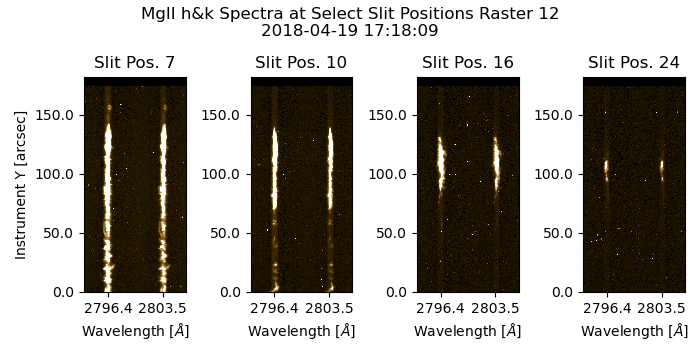}
    \includegraphics[width=0.75\linewidth]{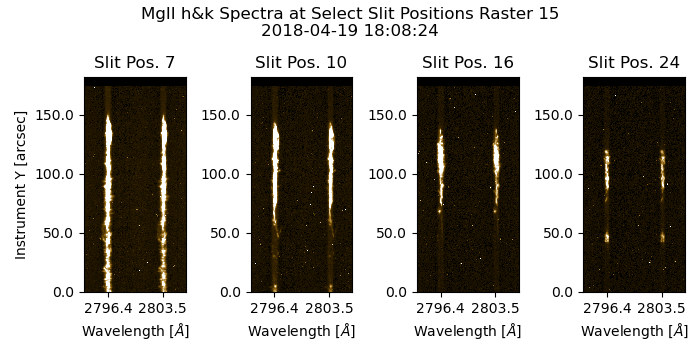}
    \caption{\ion{Mg}{ii} h\&k spectra at select slit locations. See Figure \ref{maps} for context.}
    \label{slitevo}
\end{figure*}

Some of the \ion{Mg}{ii}~h\&k spectra obtained by IRIS show complex structure. This implies that some regions of the prominence have more than one structure along the line of sight. A notable example is seen in the top of Figure \ref{slitevo} near 120\arcsec\ in slit position 7. Here we see what appears to be a single peaked profile in \ion{Mg}{ii}~h\&k with a secondary redshifted peak. This is not unique to this slit position, though this is the most extreme case. There are many areas where line profiles are composed of many intensity peaks. These lead to the appearance of more than one, or complex, looking line profiles. Figure \ref{lineprofs} shows the distribution of the spectral line shapes observed by IRIS. While the distribution of line profile types is dominated by single peaked profiles, the number of double peaked and complex profiles are not negligible -- with approximately 20\% of line profiles appearing complex.

Here we use the quantile method \citep{kerr2015, ruan2018} to determine the FWHM, line core shift, and asymmetry of the line profiles to investigate the internal structure and dynamics of the prominence. The quantile method involves calculating the cumulative distribution function (CDF) of the intensity of the individual line profiles over some wavelength range. Here, we employ a 3\AA\ window centred on the rest vacuum wavelengths of \ion{Mg}{ii} h\&k. The wavelength at which the 50\% level of the CDF is found ($\lambda_{50}$) is defined as the line core. The FWHM can be calculated by taking the difference between the wavelength at which the 88\% ($\lambda_{88}$) and the 12\% ($\lambda_{12}$) level of the CDF is found. Additionally, the asymmetry of the line profiles is calculated by,
\begin{equation}
    \text{Asymmetry}=\frac{(\lambda_{88}-\lambda_{50})-(\lambda_{50}-\lambda_{12})}{\lambda_{88}-\lambda_{12}}.
\end{equation}
The rest wavelengths of the \ion{Mg}{ii} h\&k line cores were calculated by performing quantile analysis on every pixel of the first slit position of every raster, taking $\lambda_{50}$ as the line core. This was done as the first slit position of each raster is on the solar disc, which gives us a rotation independent measure of the line cores. We use the nominal wavelength calibration provided in the data header. From this, 9360 measurements of the line core were obtained for h\&k, of which the mean was taken. The rest wavelengths for the \ion{Mg}{ii} h\&k line cores were found to be 2803.494~\AA\ and 2796.348~\AA\, respectively, whereas the rest vacuum wavelengths are 2803.53~\AA\ and 2796.35~\AA\, respectively \citep{levensmgii}. The spectral resolution of the \ion{Mg}{ii} observations were 0.051~\AA, giving these measured values an error of $\pm0.0255$\AA. The calculated wavelengths of the line cores are used here as reference wavelengths in order to effectively transform into the inertial frame of the solar limb such that its rotation is not included in the line core shift calculations.

\subsection{Line profile distribution}
\begin{figure}
    \centering
    \includegraphics[width=\linewidth]{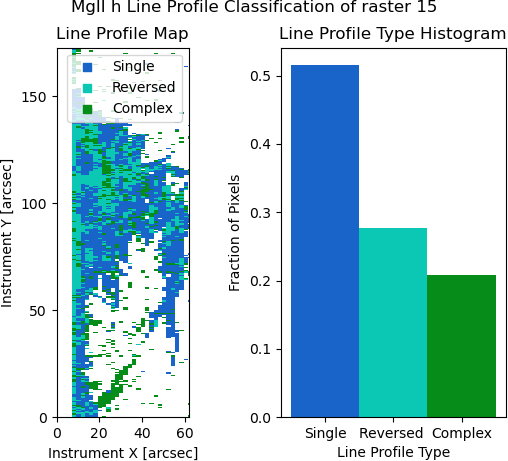}
    \caption{\ion{Mg}{ii} h line profile types seen in raster 15.}
    \label{lineprofs}  
\end{figure}
The prominence presents a variety of line profile types. These were found by using a finite difference approximation of the derivative of values above the standard deviation of the intensity of the currently considered pixel. From this, stationary  points could be counted. If two, or more than three stationary  points were found, the line profile was classified as complex. Such a complex line profile is discussed in Section~\ref{fwhm}. Figure \ref{lineprofs} shows an example of the distribution of line profile types in raster 15. 
Parts of the prominence nearest the edges of the structure are more likely to exhibit single peaked behaviour relative to those in nearer the centre. Those under motion also favour single peaked behaviour.

We observe a large arm extending from the main body of the prominence near the end of the observation that displays mainly single peaked line profiles. This arm also contains a few classic double peaked line profiles. In addition, it exhibits complex profiles that appear to be double peaks suffering from the effect further discussed in Section~\ref{vel} where the emission of h$_{2v}$ and k$_{2v}$ is greatly reduced, resulting in only h$_{2r}$ and k$_{2r}$ peaks. This is due to the intensity of h$_{2v}$ and k$_{2v}$ peaking lower than h$_{3}$ and k$_{3}$, which causes no defined peaks at these locations. Double peaks in the arm can be interpreted in one of two ways -- there is a large volume of material in the arm along the line of sight or there exists more than one strand in the composition of this arm with different line-of-sight velocity shifts. In the former case, the optical thickness of the plasma would result in a double peaked profile. However, judging by the extent of the arm in the plane-of-sky, this interpretation is unlikely. For it to be extended enough along the line of sight to cause this effect, would mean it would take the form of a sheet-like structure and we see no evidence of this morphology. A more plausible interpretation is the latter. This would lead to a measured line core shift, as if it were the former case, resulting in a lower measured velocity. This may also be another reason for the broadened wings of the line core shift distribution discussed in Section~\ref{vel}. Admittedly, this would be a rather small effect on the distribution, and the more likely cause is the effect that asymmetry has on measured Doppler velocity when using the quantile method. Highly asymmetric profiles have bias in the measure of their Doppler velocity due to the increased emission of the red or blue sides of these line profiles. This effect is discussed in detail in Section~\ref{vel}.

The double peaked line profiles of the main body of the prominence are most likely due to a large amount of material along the line of sight. The optical thickness would assist in producing the classic double peaked look of \ion{Mg}{ii} h\&k emission. The pixels classified as complex in figure \ref{lineprofs} are likely due to many different structures or threads in that pixel. These areas of complexity appear to correlate with areas of large FWHM and asymmetry, giving more evidence towards this interpretation. The south-eastern edge of the prominence displays mainly single peaked profiles, which can be attributed to a few causes. This area of the prominence is seen to contribute to many small flows, causing a loss of plasma to this area, which could lead to a lower density, or there could be less plasma along the line of sight here, or perhaps both of these factors contribute towards the observed emission.

\subsection{Line core shift}
\label{vel}
\begin{figure}
\centering
    \includegraphics[width=0.49\linewidth]{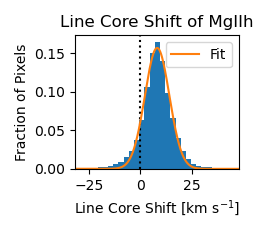}
    \includegraphics[width=0.49\linewidth]{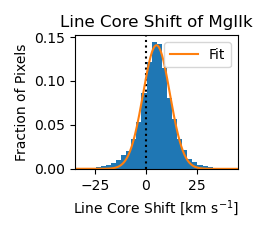}
    \caption{Line core shift histograms of both \ion{Mg}{ii} h\&k across all 18 rasters with Gaussian fits.}
    \label{doppdist}
\end{figure}
We calculate line core shifts by computing the Doppler velocity of each pixel using the 50\% quantile ($\lambda_{50}$) of each pixel as the line core. From this, we find a range of line core shifts in the prominence, with the extrema ranging from -74 to 78~km~s$^{-1}$ in \ion{Mg}{ii}~h and -72 to 85~km~s$^{-1}$ in \ion{Mg}{ii}~k. The total distribution is a Gaussian-like centred on a redshifted line core of around 8.20~km~s$^{-1}$ and 5.20~km~s$^{-1}$ with standard deviations of 5.98~km~s$^{-1}$ and 6.61~km~s$^{-1}$ in \ion{Mg}{ii} h\&k, respectively. These distributions can be seen in  Figure \ref{doppdist}. 

\ion{Mg}{ii} h\&k line shifts are an excellent diagnostic tool for the Doppler velocities seen in the structure \citep{irisii}. These recovered velocities are quite typical of past prominence observations. \cite{liggett1984} observed velocities ranging from 12 to 75~km~s$^{-1}$, and \cite{levens3d} found maximum velocities of around 60~km~s$^{-1}$ . Table 6 in the review by \cite{promreviewi} shows a range of velocities exhibited by prominences in the EUV, with an extremum of 70~km~s$^{-1}$, centred around 20~km~s$^{-1}$. 

\begin{figure}
    \includegraphics[width=0.9\linewidth]{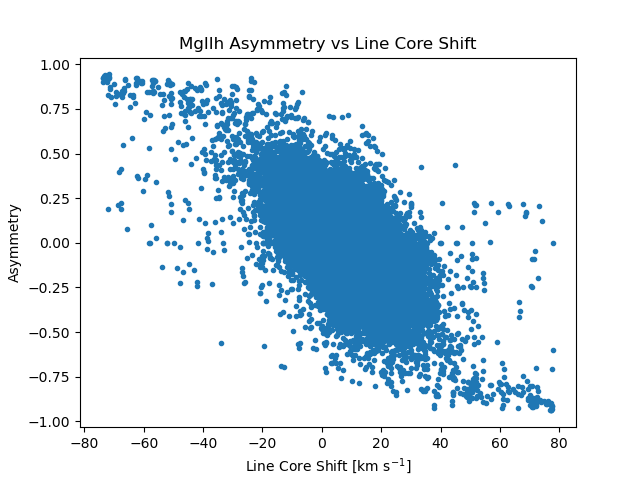}
    \caption{Scatter plot of relationship between asymmetry and measured line core shift of all post-filter \ion{Mg}{ii} h pixels.}
    \label{da}
\end{figure}

The wings of the Gaussian fits used in Figure~\ref{doppdist} do not seem to model the velocity distributions well. The cause of this shape is suggested by Figure~\ref{da}. Asymmetry and measured Doppler velocity appear to be anti-correlated, plateauing at high measured velocities.
The root of this trend lies in the optical thickness of the lines. At rest, the line cores of \ion{Mg}{ii} h\&k  are optically thick, but the wings are optically thin. Therefore, line profiles of high redshift will display a negative (red) asymmetry due to h$_{2v}$ and k$_{2v}$ being shifted into the more optically thick area of the line. This leads to the quantile method measuring a larger-than-true velocity due to the absorbed h$_{2v}$ and k$_{2v}$ giving more credence to the intensity of h$_{2r}$ and k$_{2r}$. This effect is amplified by h$_{2r}$ and k$_{2r}$ moving more into the optically thin regime of the wings. The opposite is also true for the blueshifted profiles that see absorption and enhancement of h$_{2r}$ and k$_{2r}$ and  h$_{2v}$ and k$_{2v}$, respectively. The extensions of the wings of the Doppler velocities are likely due to this effect. It is therefore sensible to assume these high velocities are measured to be larger than in reality, and the true distribution may still be of a Gaussian shape. The velocities of areas where the asymmetry is much greater than zero are likely unreliable, and only those where the asymmetry is approximately 0, can be considered reliable.

\subsection{Line widths}
\label{fwhm}
\begin{figure*}
    \centering
    \resizebox{\hsize}{!}
    {\includegraphics{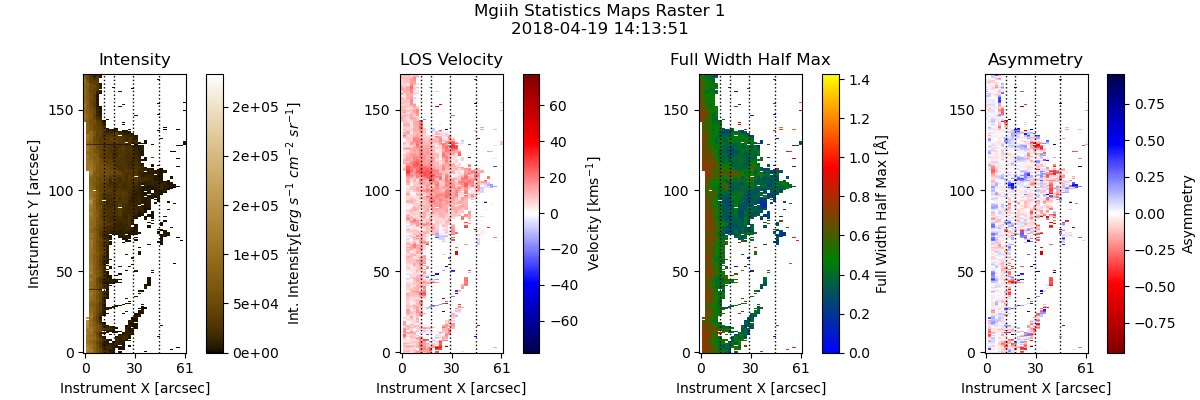}}
    \resizebox{\hsize}{!}
    {\includegraphics{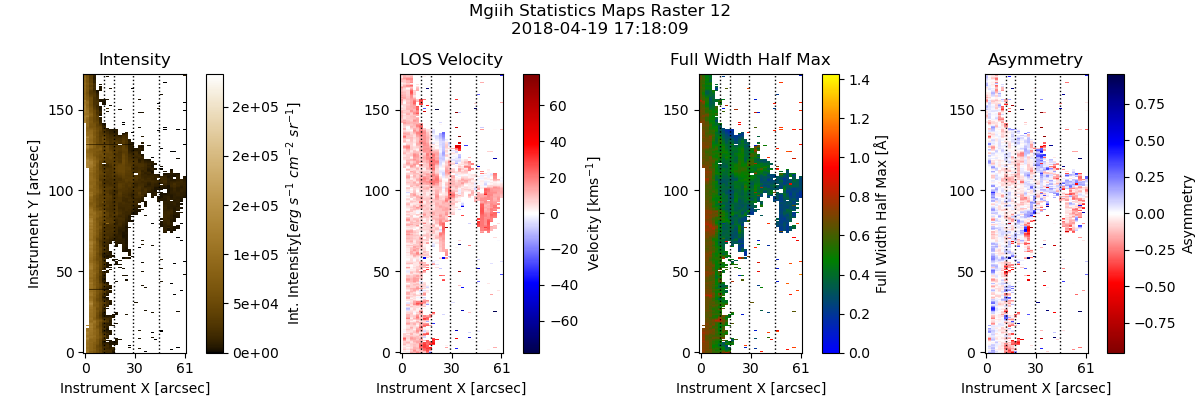}}
    \resizebox{\hsize}{!}
    {\includegraphics{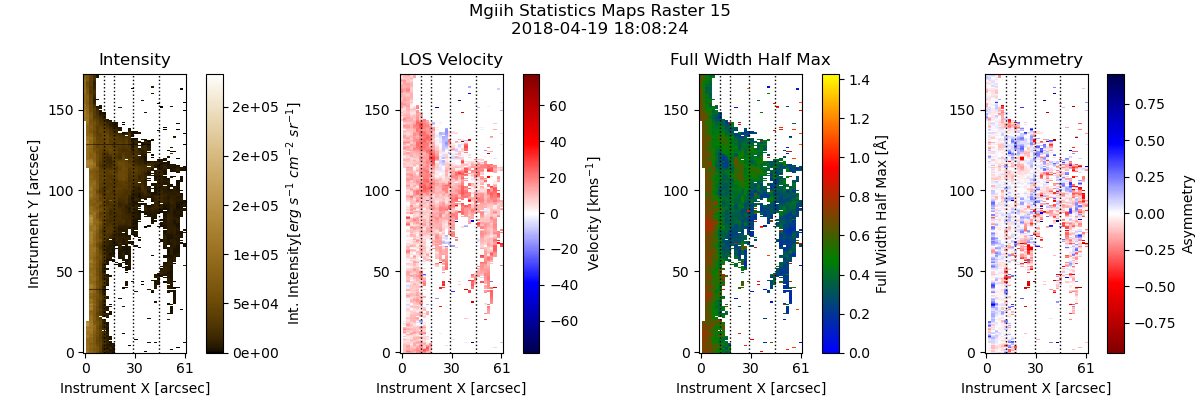}}
    \caption{\ion{Mg}{ii} h statistics maps of the three main stages of the prominence observation calculated via the quantile method. In the first raster we see a rather static structure. In raster 12, the rise that eventually leads into a flow begins to form. In raster 15, the flow seen in raster 12 is beginning. The times associated with these plots are the time at the beginning of the associated raster. The dashed lines represent the slit positions in Figure~\ref{slitevo}.  \ion{Mg}{ii} k produces similar plots and these can be seen in Figure~\ref{kmaps}. An animated version of this figure~is available online.}
    \label{maps}
\end{figure*}

Here, we use the difference between the 88\% and 12\% quantiles ($\lambda_{88}-\lambda_{12}$) to measure the FWHM, as in \cite{ruan2018}. If you assume that the line profiles are of a Gaussian shape, the difference between the 88\% and 12\% quantiles is the FWHM.
FWHM maps clearly separate the prominence (and spicules) from the solar disc. They also separate the prominence and solar disc from the background corona as discussed in Section~\ref{irisobs}. These FWHM maps allow for the location of flows and structures not easily seen in intensity maps. The FWHM of the lines profiles within the prominence are relatively small, ranging from around 0.1~\AA\ to 0.6~\AA, whereas the FWHM within the solar disc have values of around 0.8~\AA\ to 1~\AA. In the first raster (14:13UTC; see Figure~\ref{maps}), we see a structure of large FWHM, in comparison to the same location in later rasters, around (10\arcsec, 110\arcsec). This structure is seen in both \ion{Mg}{ii} h\&k , suggesting that this is a physical phenomena. On further inspection, the spectra in both \ion{Mg}{ii} h\&k, in this area, show two distinct line profiles, one of which is highly redshifted. The shape and location of this structure seems very similar to that of the pillar-like barb easily seen in 171~\AA. One of the drawbacks of the quantile method here is that it cannot distinguish between line profiles. If a pixel displays more than one line profile, or one of a more complex nature, the algorithm is still able to produce a number for this, albeit a large one. This can then be used as a proxy for whether a spatial pixel contains more than one line profile.

\subsection{Asymmetry}
Asymmetry is most apparent in the extended arm protruding from the main prominence body as seen in raster 15 (see Figure~\ref{maps}). Most emission in this area is seen to have asymmetry in the range [0,-1). However, these line profiles may just have too low of an intensity, which gives too much credence to the noise in the quantile analysis. Many of these profiles do not appear asymmetric, just highly Doppler shifted. This can be seen in Figure~\ref{maps} as the areas of high redshift appear to correlate to those of non-negligible asymmetry and low integrated intensity. However, there does appear to be an anticorrelation between the line core shift and the asymmetry as can be seen in Figure~\ref{da} The only structure that appears to break this trend, is the structure of large FWHM discussed in Section~\ref{fwhm}. Its large asymmetry is likely due to the two line profiles seen in this area, leading to a erroneous measurement of asymmetry. This shows that asymmetry calculated via the quantile method, when combined with its complimentary FWHM, is a better proxy for whether a pixel contains more than one line profile than either diagnostic on its own.

\begin{table}
\begin{tabular}{ccc}
\hline \hline
Parameter & Unit           & Value                                                                                                             \\ \hline\hline
$T_{\text{cen}}$   & K              & \begin{tabular}[c]{@{}c@{}}6000, 8000, 10000, 15000, \\ 20000, 25000, 30000, \\ 35000, 40000\end{tabular}         \\
$T_{\text{tr}}$    & K              & 100000 \\
$p_{\text{cen}}$   & dyne cm$^{-2}$ & 0.01, 0.02, 0.05, 0.1, 0.2, 0.5, 1                                                                                \\
$p_{\text{tr}}$     & dyne cm$^{-2}$ & 0.01                                                                                \\
Slab Width & km             & 200 -- 124100                                                                                                     \\
M         & g cm$^{-2}$    & $3.7\times10^{-8}$--$5.1\times10^{-4}$                                                                            \\
$v_T$ & km~s$^{-1}$ & 5 \\
$H$ & km & 1000 \\ 
$\gamma$  &                & 0, 2, 5, 10  \\ \hline
\end{tabular} 
\caption{Parameters of the models.}
\label{modeltable}
\end{table}

\section{Model comparisons}
\label{models}
\cite{levensmgii} discussed an extension to the one-dimensional radiative transfer code developed by \cite{GHV}, PROM, such that it was able to produce \ion{Mg}{ii} line profiles with both isobaric and isothermal atmospheres, and atmospheres containing a PCTR. The latter atmosphere was itself added as an extension by \cite{promctr}. This \ion{Mg}{ii} extension is similar to that discussed by \cite{zhangcmepromiris}, but independently developed. In this model, prominences are represented by a one-dimensional semi-infinite plane-parallel slab perpendicular to the solar surface. This block is illuminated on both sides by isotropic radiation from the solar disc. Using this extension developed by \cite{levensmgii}, we produced a total of 1007 \ion{Mg}{ii} model profiles, 252 of which are isothermal and isobaric, where the remaining 755 include a PCTR (one of the original set of 756 did not converge). The parameters of these atmospheres can be seen in Table~\ref{modeltable}. $T_{\text{cen}}$ and $p_{\text{cen}}$ are the central temperature and pressure, respectively. $T_{\text{tr}}$ and $p_{\text{tr}}$ are the temperature and pressure at the edge of the slab, respectively. Slab width is the width of the slab. $M$ is the column mass. $v_T$ is the microturbulent velocity inside the prominence. $H$ is the height above the solar surface. $\gamma$ is a dimensionless number that dictates the extent of the PCTR. A $\gamma$ value of 0 indicates the model is isothermal and isobaric -- without a PCTR. For isothermal and isobaric models, $T_{\text{tr}}=T_{\text{cen}}$, and $p_{\text{tr}}=p_{\text{cen}}$. For non-zero values of $\gamma$, the lower the value of $\gamma$ the more extended the PCTR is.  All of the above combinations amount to 1008 models; however, one model did not converge, so we only consider the 1007 that did.
The PCTR models used here adopt the same parametric description of the PCTR as in \cite{ah99}. The pressure profile derives from magnetohydrostatic equilibrium and is given as a function of the column mass \(m\) by
\[
p(m)=4p_c\frac{m}{M}\left(1-\frac{m}{M}\right)+p_{\text{tr}} \ ,
\]
where \(p_c\) is the difference in pressure between slab centre and slab edge  (\(p_{\text{cen}}=p_c+p_{\text{tr}}\)), and $M$ is the total column mass.
The temperature profile is taken to be 
\begin{equation}
    T(m)=T_{\text{cen}}+(T_{\text{tr}}-T_{\text{cen}})\left(1-4\frac{m}{M}\left(1-\frac{m}{M}\right)\right)^\gamma \ ,
    \label{eq:gamma}
\end{equation}

Using these models we are able to estimate the internal plasma properties of the prominence. The models contain the five main \ion{Mg}{ii} lines, here, we focus on the h\&k lines in particular. The computed line profiles are symmetrical around their respective rest wavelengths due to the omission of radial velocities, which would introduce the effects of Doppler dimming and/or Doppler brightening \citep{1970hyder, 2007labrosse}.

In order to compare these model line profiles with the observations, we developed a `rolling root mean squared (RMS)' procedure  in Python. This works by first downgrading, through linear interpolation, the resolution of the model line profiles to that of our IRIS observations. The models and observations have a resolution of 3$\times10^{-3}$\AA\ and 51$\times10^{-3}$\AA, respectively. Once downgraded, a spectral window of 3\AA\, centred first on the rest wavelength of \ion{Mg}{ii} h, then \ion{Mg}{ii} k, is considered. This spectral window is considerably larger than the model range, and so, the models are padded with zeroes to match the size of the spectral window. 
\begin{equation}
    \text{RMS}=\sqrt{\frac{1}{n}\sum_n\left(\text{data}-\text{model}\right)^2}.
    \label{rms}
\end{equation}
We define RMS as seen in equation \ref{rms}, where $n$ is the number of wavelength points.
The RMS is first measured between the downgraded model at beginning of this spectral window and the observations. The model is then shifted one pixel to the right, and the RMS is measured again. This process repeats until the model reaches the end of the spectral window. The lowest measured RMS from this `roll' is recorded, along with the pixel shift required to achieve this. The next model in the set is then tested. This process is repeated until all of the models have been tested. This is run for both h\&k. The RMS values for h\&k are then summed, and the model that produces the lowest overall RMS sum is then selected as the model indicative of the plasma parameters if the RMS is lower than a certain threshold. RMS values that are larger than that threshold indicate that the computed and observed line profiles are too different, and that no model could be found to satisfactorily reproduce the observed line profiles.

By `rolling' our models through the spectral window, this gives us a measure of the wavelength of the line core (from the recorded pixel shift required to minimise the RMS) and allows us to calculate a Doppler velocity. The rolling RMS (rRMS) procedure is also able to measure pixel shift with sub-pixel precision. However, the run time of the procedure scales linearly with the number of sub-pixels requested. This works identically to without sub-pixel precision, but before the rolling begins, the models and data are interpolated up to match the sub-pixel precision requested.

\subsection{Results}
\begin{figure*}
    \centering
    \resizebox{\hsize}{!}
    {\includegraphics[width=0.49\linewidth]{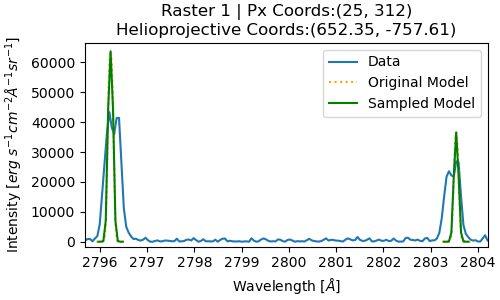}
    \includegraphics[width=0.49\linewidth]{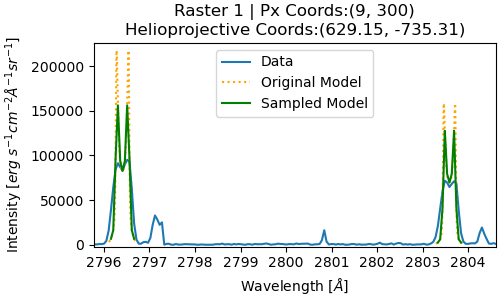}
    }
    \resizebox{\hsize}{!}
    {\includegraphics[width=0.49\linewidth]{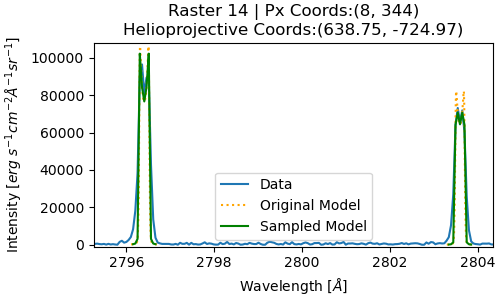}
    \includegraphics[width=0.49\linewidth]{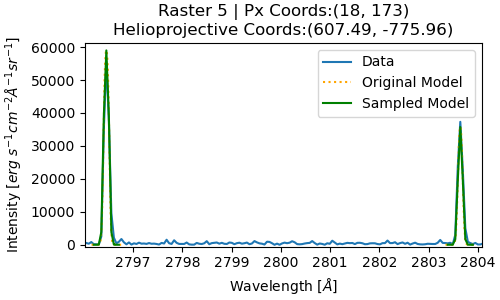}
    }
    \caption{Four of the twelve random line profiles where a good fit is found for six and an unsatisfactory fit is found for the other six. The top two panels are the unsatisfactory fits, and the bottom two panels are the satisfactory fits.}
    \label{fig:egpx}
\end{figure*}

\begin{figure}
    \centering
    \includegraphics[width=0.49\linewidth]{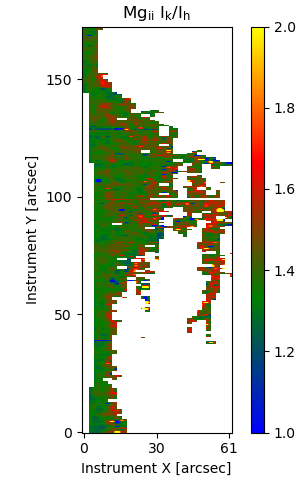}
    \includegraphics[width=0.49\linewidth]{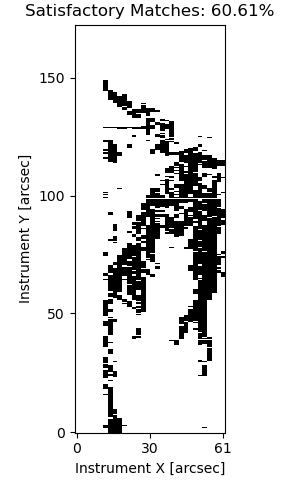}
    \caption{Comparison of k/h ratio and satisfactory fits. This demonstrates the weak correlation between high k/h ratio and satisfactory fits. (left) k/h ratio of raster 15. (right) Satisfactory fits of raster 15. Raster 15 was taken between 18:08 and 18:25 UTC.}
    \label{fig:khandgoods}
\end{figure}

After the procedure was run, twelve line profiles were selected at random to illustrate where the procedure found good matches and where it did not (see Figure~\ref{fig:egpx}). 
The goodness of these fits were judged by eye. The measured RMS of these twelve line profiles were used to arrive at a cut off value above which matches would be considered unsatisfactory. This value was determined to be 15000, concluding that 49\%, 35617/72536, of the matches were satisfactory. Figure \ref{fig:khandgoods} shows that areas of higher k/h ratio weakly correlate with areas that give a satisfactory fit. High values of the k/h ratio are related to areas of lower optical thickness. Therefore, this demonstrates that areas of lower optical thickness lead to better matches with the models. This could also be due to the one-dimensional nature of the code as the satisfactory fits seem to also correlate with areas of single peaked line profiles, as seen in Figure~\ref{lineprofs}. These values for the k/h ratio appear to be consistent with \cite{aliss2018} where they found a k/h ratio of 1.22 on approach to the limb and values up to 1.96 in the spicules.

\begin{figure}
    \centering
    \includegraphics[width=.49\linewidth]{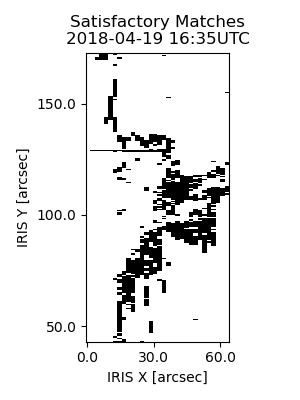}
    \includegraphics[width=.49\linewidth]{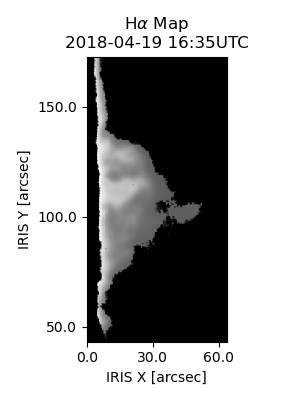}
    \includegraphics[width=.6\linewidth]{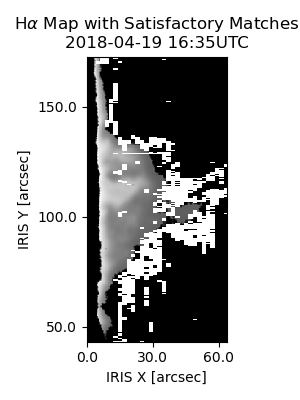}
    \caption{MSDP H$\alpha$ integrated intensity with corresponding satisfactory matches. 
    \textit{Top Left}: Raster 9 satisfactory matches. \textit{Top Right}: MSDP H$\alpha$ integrated intensity. \textit{Bottom}: H$\alpha$ integrated intensity map from MSDP with the satisfactory matches of raster 9 overlaid in white.}
    \label{fig:msdp}
\end{figure}

Figure \ref{fig:msdp} shows that in the most part, areas in which we see H$\alpha$ emission in the prominence correspond to where computed \ion{Mg}{ii} line profiles do not satisfactorily match the observed ones. The observed H$\alpha$ line profiles are generally complex in shape and much wider than the models produce. This issue is seen when attempting direct one-dimensional forward modelling of H$\alpha$ prominences. For example, \cite{ruan2019} demonstrated that areas with wide H$\alpha$ line profiles can be better fit by models with high microturbulent velocities. Whether this is also true for \ion{Mg}{ii} line profiles is currently unclear.
Figure~\ref{fig:temporal} presents the temporal evolution of the mean pressure and mean temperature as derived from the comparison between computed and observed \ion{Mg}{ii}~h\&k line profiles, while Fig.~\ref{fig:temp_pres} shows these parameters for the pixels in rasters 1, 12, and 15 recovered by the procedure. Figure~\ref{fig:gamm_cmass} presents diagnostic maps for electron density, ionisation degree, $\gamma$ (i.e. temperature gradient), and column mass for raster 15. We stress that the plasma parameters obtained with this method are only reliable where our RMS values are low enough, as indicated with the satisfactory matches in the right panel of Figure~\ref{fig:khandgoods}.

\subsubsection{Pressure}
\begin{figure}
    \centering
    \includegraphics[width=\linewidth]{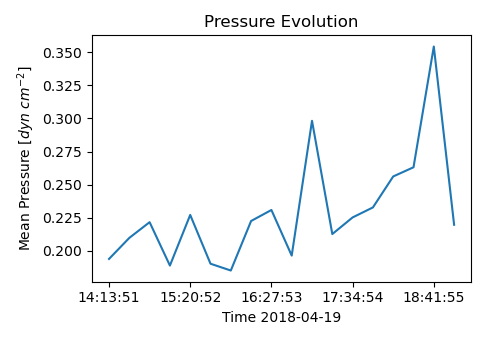}
    \includegraphics[width=\linewidth]{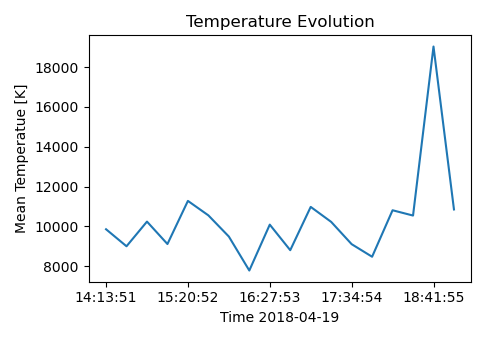}
    \caption{Temporal evolution of (top) mean pressure and (bottom) mean temperature. These plots only include the satisfactory matches. The large spike near the end is due to interference seen in half of the slit positions of that raster. This caused a lot of the previously lower temperature regions to be filtered out, and therefore only the higher temperature areas are included in the calculation of the mean. A similar (but less extreme) situation occurred where the second highest peak is seen in the plot of pressure; however, the mean temperature plot seems unaffected by this.}
    \label{fig:temporal}
\end{figure}

The mean pressure appears to remain stable during the IRIS observation, with the mean pressure fluctuating on average between 0.18 and 0.26 dyne cm$^{-2}$ (Fig.~\ref{fig:temporal}). The pressures of 1~dyn~cm$^{-2}$ found  towards the core of the structure correlate closely with the unsatisfactory fits (Fig.~\ref{fig:temp_pres}). This high pressure is at the upper limit of that outlined in Table~1 of \cite{promreviewi}. These selected models could be explained by the procedure finding that broadened profiles produce a smaller RMS for line profiles that exhibit a complicated structure due to the large volume of the prominence to travel through (a larger optical depth) coupled with many different threads along the line of sight as seen in this area of the prominence. Pressure near the outer edges of the prominence is, in general, lower as we move through the PCTR towards the corona where the gas pressure is lower \citep{aschwandencorona}.

\subsubsection{Temperature}
\begin{figure}
    \centering
    \includegraphics[width=0.47\linewidth]{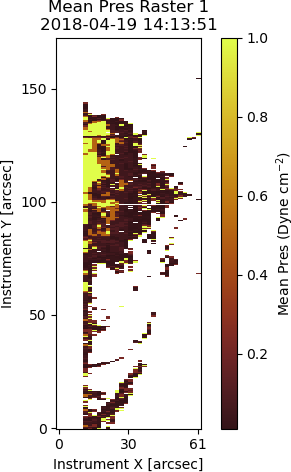}\hspace{0.05cm}
    \includegraphics[width=0.51\linewidth]{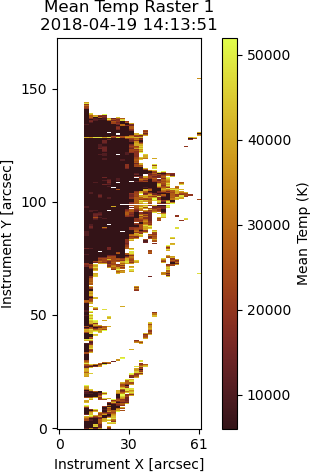}
    \includegraphics[width=0.47\linewidth]{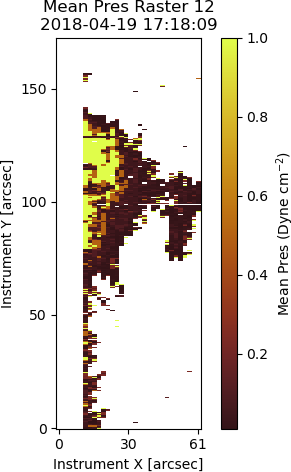}\hspace{0.05cm}
    \includegraphics[width=0.51\linewidth]{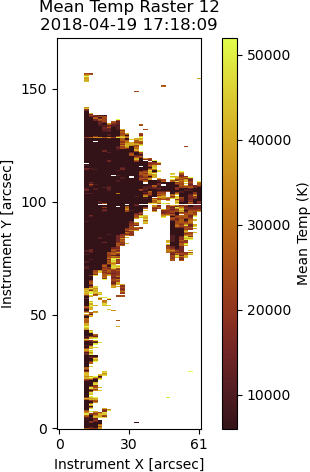}
    \includegraphics[width=0.47\linewidth]{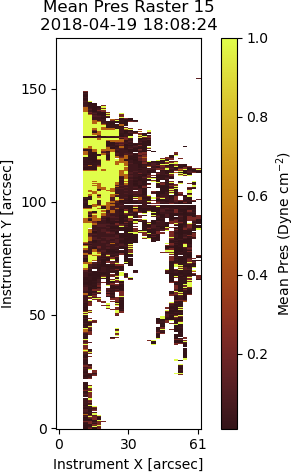}\hspace{0.05cm}
    \includegraphics[width=0.51\linewidth]{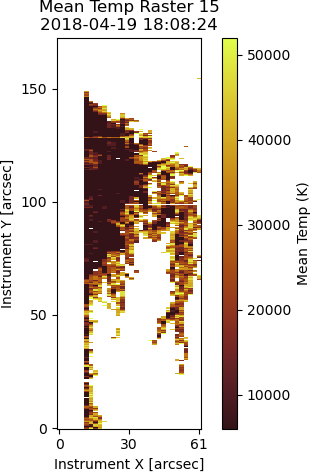}
    \caption{Evolution of mean temperature and pressure in the prominence from the model comparisons with rasters 1, 12, and 15.}
    \label{fig:temp_pres}
\end{figure}
\begin{figure}
    \centering
    \includegraphics[width=0.50\linewidth]{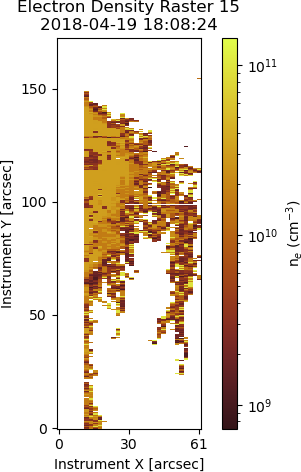}
    \includegraphics[width=0.49\linewidth]{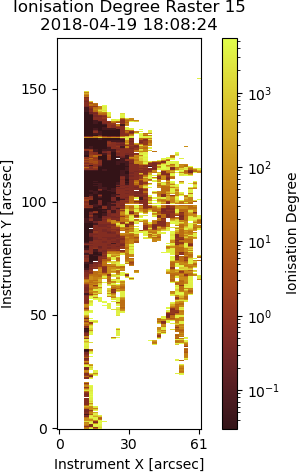}\\
    \includegraphics[width=0.46\linewidth]{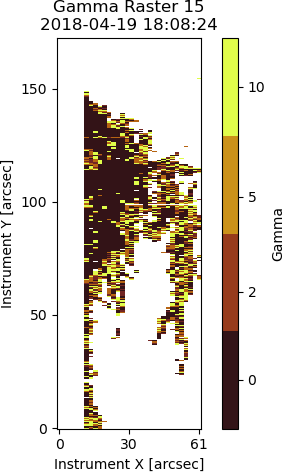}\hspace{0.05cm}
    \includegraphics[width=0.52\linewidth]{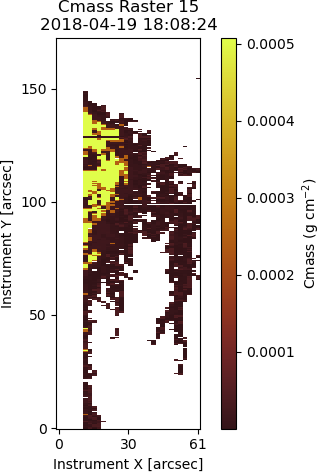}
    \caption{Output diagnostics for raster 15: (top left) mean electron density; (top right) ionisation degree; (bottom left) column mass; (bottom right) $\gamma$ values. Raster 15 was taken between 18:08 and 18:25 UTC.}
    \label{fig:gamm_cmass}
\end{figure}

The mean temperature, like pressure, also appears stable during the observation, with the mean temperature staying on average between $7~800$K and $11~500$K (see Figure~\ref{fig:temporal}). This is consistent with expected results. The formation height of \ion{Mg}{ii} h\&k ranges from the upper photosphere to the upper chromosphere \citep{iris}. However, in a prominence, we expect a range of chromospheric to PCTR temperatures. The temperature of the prominence increases towards the outer edges where one would expect to  see the PCTR in the plane of sky (Fig.~\ref{fig:temp_pres}). Figure~\ref{fig:gamm_cmass} further reveals higher values nearer to the edges of the prominence. The arm seen near the end of the observation appears to be a tenuous and hot structure reaching upper chromospheric temperatures of a few $10^4$~K. 

Equation \ref{eq:gamma} describes the relationship between \(\gamma\) and the temperature profile $T(m)$ (a function of column mass).
In Table~\ref{modeltable}, models where $\gamma=0$ in fact correspond to isothermal and isobaric models without a PCTR.
Many of the satisfactory fits having $\gamma\neq0$ shows that the PCTR extension to PROM by \cite{promctr} implemented into the \ion{Mg}{ii} extension by \cite{levensmgii} is important to obtain a better agreement to the data than isothermal and isobaric models. 
Our mean temperatures are greater than temperatures obtained by \cite{zhangcmepromiris}. However, the authors of that study employed isobaric and isothermal models whereas our models include a PCTR, as mentioned before. This allows our inferred mean temperatures to be much greater. Areas where $\gamma=0$, indicating isothermal models,  correlate with areas of lower temperature, which agrees with their study.
In future, a finer grid of $\gamma$ values could help with a better resolution of the temperature gradient of the PCTR.

\subsubsection{Column mass}
\begin{figure}
    \includegraphics[width=.9\linewidth]{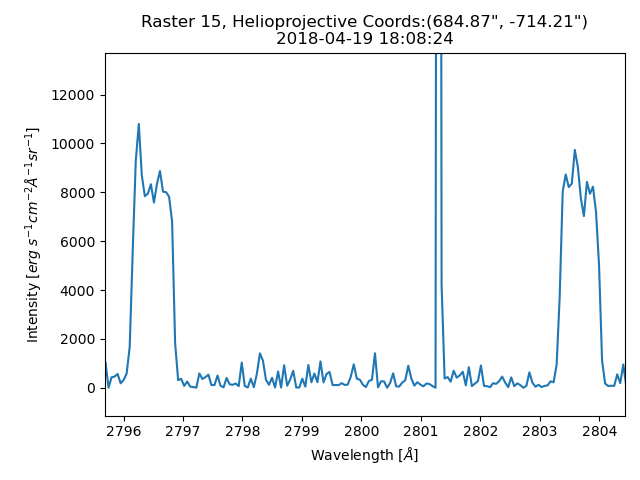}
    \caption{Line profiles of \ion{Mg}{ii}~h\&k in the region where unsatisfactory fits are found. These line profiles can be explained by many threads being seen along the line of sight moving at different velocities. The large spike around 2801.3\AA\ is noise.}
    \label{fig:complex_prof}
\end{figure}

The mass along the line of sight conveys that the extensions of the prominence have comparatively low mass (Fig.~\ref{fig:gamm_cmass}). The region where unsatisfactory fits are found is much higher density than the surrounding material. While these fits may be unsatisfactory, we can infer from the line profiles of that area that there are several threads along the line of sight that contribute to the line profiles. This includes profiles such as those shown in Figure~\ref{fig:complex_prof}. The unsatisfactory fits here are due to the single slab nature of the models, which does not account for several threads along the line of sight. A single slab model can only apply, when there are several threads, if the threads are comoving, which would result in a single line profile. The arm seen near the end of the IRIS observations appears to be comparatively low mass. This explains the single peaked line profiles observed in the arm. The intensity of the central reversal is anticorrelated with formation height due to optical depth \citep{irisii}. This reinforces the finding of low column mass not just in the arm, but where single peaked line profiles are observed. 

\subsubsection{Electron density}
The majority of the prominence appears to have an electron density of less than 0.6$\times10^{11}$cm$^{-3}$, with the exception of a few places where it is greater than 1$\times10^{11}$cm$^{-3}$ (Fig.~\ref{fig:gamm_cmass}). The arm appears to be of generally comparatively low electron density. Even though it is of a comparatively high temperature compared to the rest of the prominence, its low pressure can explain the low electron density found. 

\subsubsection{Ionisation degree}
\begin{figure}
    \centering
    \includegraphics[width=.8\linewidth]{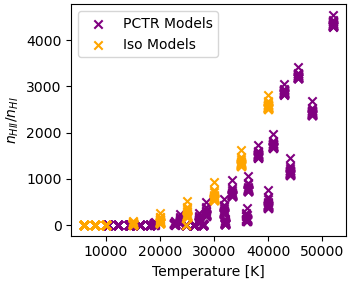}
    \includegraphics[width=.8\linewidth]{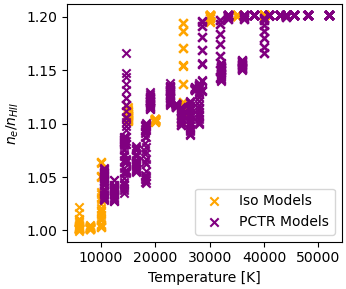}
    \caption{Exponential increase of ionisation degree above 30~000K and breakdown of the assumption that $n_{\text{HII}}\approx n_e$ at high temperatures. Top: Ionisation degree defined by $n_\text{HII}/n_\text{HI}$. Bottom: Excess electrons. Number of free electrons per \ion{H}{II}. This illustrates that the commonly used approximation for ionisation degree will overestimate the value of $n_\text{HII}$ at high temperatures.}
    \label{ionisation_degrees}
\end{figure}
Ionisation degree is defined by the mean number density of \ion{H}{II} ($n_\text{HII}$) to \ion{H}{I} ($n_\text{HI}$) \citep{hanssen1995}. Past works \citep[such as][]{zhangcmepromiris} assume that the mean number of electrons ($n_\text{e}$) is equal to the mean number of protons ($n_\text{p}$), and further that the mean number of protons is equal to the mean number density of \ion{H}{II} and that the mean number density of \ion{H}{I} is dominated by the mean number density of ground state Hydrogen ($n_\text{H0}$) then,
\begin{equation}
    \frac{n_\text{HII}}{n_\text{HI}}\approx\frac{n_\text{e}}{n_\text{H0}}.
\end{equation}
However, the assumption that $n_\text{e}\approx n_\text{p}\approx n_\text{HII}$ does not hold with PROM because not all of the electrons are liberated from \ion{H}{I}. There is a fixed contribution of electrons from other species, such as helium, that contribute to the total number of electrons (named excess electrons from here on). This increase is most apparent at high temperatures when the number density of \ion{H}{II} and excess electrons increase dramatically. Figure \ref{ionisation_degrees} shows the behaviour of ionisation degree and excess electrons with increasing temperature. At temperatures above 30~000K, these species are responsible for a 20\% increase in the total number of electrons (see Figure~\ref{ionisation_degrees} bottom). So here we focus on the definition from \cite{hanssen1995}.
Some sections of the prominence are found to have a very large ionisation degree when compared to past studies. \cite{vial1998} summarised that the ionisation degree in a prominence does not change much with temperature, given that the density is low. This was based on studies by \cite{GHV} and \cite{HGV}. However, the highest temperature considered in these studies is 15~000K. Figure \ref{ionisation_degrees} shows that the ionisation degree is fairly constant up until 15~000K, consistent with these studies, but then rises sharply at higher temperatures. Additionally, at 15~000K, the excess electrons are responsible for only a roughly 7\% increase in the number of free electrons (as in Figure~\ref{ionisation_degrees} bottom).

The areas where a high ionisation degree is found (see top right panel of Fig.~\ref{fig:gamm_cmass}) correlate strongly with areas of high mean temperature and non-zero values of gamma. The model that yields the highest ionisation degree has a central temperature of 40~000~K and a surface temperature of $10^5$~K. 

\subsection{Recovered line core shift}
\begin{figure}
    \includegraphics[width=.46\linewidth]{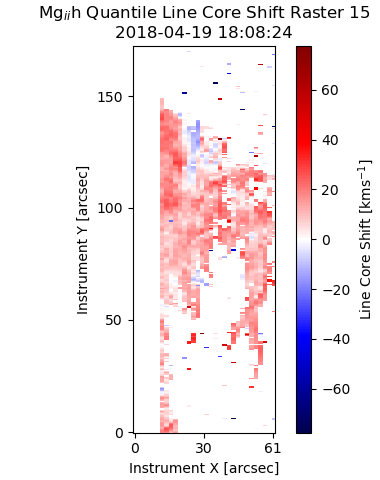}
    \includegraphics[width=.46\linewidth]{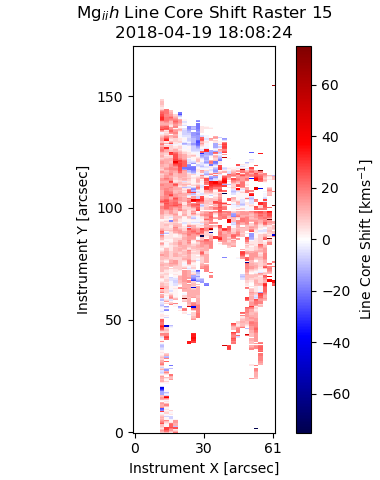}
    \caption{Comparison of line core shift recovered by quantile method and rolling RMS procedure. \textit{Left}: Line core shift recovered by quantile method. \textit{Right}: Line core shift recovered by rolling RMS procedure.}
    \label{fig:dopps}
\end{figure}

Due to the rolling aspect of the procedure, we recover a measure of the line core shift. The resolution of the line core shifts from the rolling RMS procedure is 2.7~km~s$^{-1}$. This comes from the minimum measurable line core shift that is equal to the wavelength resolution of the instrument, divided by the number of sub-pixels plus 1. Figure \ref{fig:dopps} shows the line core shift measured by the quantile method and measured by the rolling RMS procedure. These plots qualitatively agree. With the other diagnostics, there is a degree of scepticism surrounding regions where unsatisfactory fits were found. However, while those models do not represent the underlying conditions of the prominence in those regions, the fit found by the procedure will still give a good measure of line core shift. The lowest RMS will be found where the model profile is centred on the data profile, regardless of how good the fit is. Therefore, we can reasonably trust the line core shifts found by the rolling RMS procedure, even in regions where unsatisfactory fits are found.
\begin{figure}
    \includegraphics[width=0.49\linewidth]{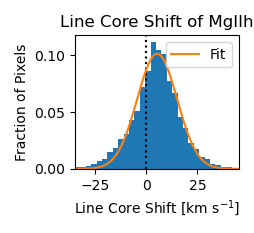}
    \includegraphics[width=0.49\linewidth]{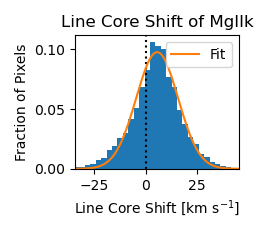}
    \caption{Histograms of line core shift recovered by the rolling RMS procedure of both \ion{Mg}{ii} h\&k across all 18 rasters with Gaussian fits. These include the unsatisfactory matches and have the filters discussed in Section~\ref{irisobs} applied.}
    \label{doppdistfound}
\end{figure}
Figure~\ref{doppdistfound} shows a similar plot to that of Figure~\ref{doppdist}, but for the line core shift recovered by the rolling RMS procedure. Fewer bins are used in the histogram of Figure~\ref{doppdistfound} than in Figure~\ref{doppdist} as the velocity resolution of Figure~\ref{doppdist} is much greater than that of Figure~\ref{doppdistfound}. The distributions in Figure~\ref{doppdistfound} display a Gaussian-like distribution, similar to what is found in Figure~\ref{doppdist}, centred on a redshifted line core of around 5.3~km~s$^{-1}$ and 5.6~km~s$^{-1}$ with standard deviations of 9.8~km~s$^{-1}$ and 10.2~km~s$^{-1}$ in \ion{Mg}{ii} h\&k respectively. Due to the limited resolution of the line core shifts recovered by the rolling RMS procedure, an estimate of the uncertainty on these values is given as $\pm$ 1.35~km~s$^{-1}$, half of the velocity resolution of the recovered line core shift. The central line core shift of \ion{Mg}{ii}~k found by the quantile method, is within the uncertainty of the central line core of \ion{Mg}{ii}~k recovered by the rolling RMS procedure. However, the central line core shift of \ion{Mg}{ii}~h does not give the same result. This could be due to the introduced asymmetry bias discussed in Section~\ref{vel}.

\section{Conclusions}
The prominence observed on the south-western solar disc on 19 April 2018 exhibited many dynamic flows seen both in AIA and IRIS. These flows were best seen in chromospheric filters. Coronal filters showed the barb of the prominence, and when it was sufficiently raised in altitude, an outline of the PCTR was also seen.

From the line core shift measurements via the quantile method, we find that the prominence is dominated by redshifted flows relative to the solar disc. Large values of asymmetry can be attributed to the plasma moving fast enough such that part of the line is emitted in the optically thin regime, closer to the wings of \ion{Mg}{ii}~h\&k emission from plasma at rest. This resulting large asymmetry biases the line core shift calculations to be redder or bluer depending on the direction that the plasma is moving and as such measures higher magnitudes for line core shift than would be expected if it were emitted closer to the rest line core. 
Therefore, using this method, the distribution of line core shifts may deviate from the expected Gaussian shape in the wings of the line core shift distribution. The prominence exhibited a wide range of line profile shapes, a majority of which were single peaked, due to the low column mass exhibited by the majority of the prominence.

Through comparisons with one-dimensional radiative transfer models, we retrieve internal plasma diagnostics of the prominence. The prominence was found to display a wide range of pressures,  ranging from 0.01--0.5 dyne$~$cm$^{-2}$. Mean temperatures ranged from 6~000--50~000K. These results included models with a PCTR, so while these temperatures may seem higher than classically thought of in a prominence, these suggest the line of sight intercepts the prominence material and much of the PCTR. We also note that the high mean temperature values returned by the models fitting the \ion{Mg}{ii} lines are found in areas where there is no detectable H$\alpha$ emission. This demonstrates the importance of the inclusion of the PCTR in prominence modelling. The column mass ranges from 3.7$\times10^{-8}$--5$\times10^{-4}$~g cm$^{-2}$.

The line core shifts recovered from our \ion{Mg}{ii} line profile comparison procedure agree well with the measured line core shifts for \ion{Mg}{ii}~k. For \ion{Mg}{ii}~h, they are of the same order of magnitude. This demonstrates that the unsatisfactory fits are not a result of the rolling RMS procedure but are a consequence of the limitations of the one-dimensional radiative transfer models. Models with multiple threads \citep{2016Rodger, 2007Gunar, 2008Gunar} are likely to produce line profiles in better agreement with the observations. 

\begin{acknowledgements}
    AWP acknowledges financial support from STFC via grant ST/S505390/1. NL acknowledges support from STFC grants ST/P000533/1 and ST/T000422/1. The authors would like to thank Dr G. S. Kerr for their assistance in the deconvolution of the IRIS spectral data, and Dr P. Mein for supplying the H$\alpha$ data from MSDP. IRIS is a NASA small explorer mission developed and operated by LMSAL with mission operations executed at NASA Ames Research Center and major contributions to downlink communications funded by ESA and the Norwegian Space Centre. AIA data courtesy of NASA/SDO and the AIA, EVE, and HMI science teams. STEREO Full-disk EUVI images are supplied courtesy of the STEREO Sun Earth Connection Coronal and Heliospheric Investigation (SECCHI) team. Hinode is a Japanese mission developed and launched by ISAS/JAXA, collaborating with NAOJ as a domestic partner, NASA and UKSA as international partners. Scientific operation of the Hinode mission is conducted by the Hinode science team organized at ISAS/JAXA. This team mainly consists of scientists from institutes in the partner countries. Support for the post-launch operation is provided by JAXA and NAOJ (Japan), UKSA (U.K.), NASA, ESA, and NSC (Norway). This research used version 2.0.3 of the SunPy open source software package \citep{sunwhy}, version 3.3.2 of Matplotlib \citep{matplotlib}, version 1.20.2 of NumPy \citep{numpy}, version 1.5.3 of SciPy \citep{scipy}, and version 4.0.1 of Astropy, (http://www.astropy.org) a community-developed core Python package for Astronomy \citep{astropy:2013, astropy:2018}
\end{acknowledgements}

%
%

\bibliography{bibtex}

\begin{thebibliography}{48}
\expandafter\ifx\csname natexlab\endcsname\relax\def\natexlab#1{#1}\fi

\bibitem[{{Alissandrakis} {et~al.}(2018){Alissandrakis}, {Vial}, {Koukras},
  {Buchlin}, \& {Chane-Yook}}]{aliss2018}
{Alissandrakis}, C.~E., {Vial}, J.~C., {Koukras}, A., {Buchlin}, E., \&
  {Chane-Yook}, M. 2018, \solphys, 293, 20

\bibitem[{{Anzer} \& {Heinzel}(1999)}]{ah99}
{Anzer}, U. \& {Heinzel}, P. 1999, \aap, 349, 974

\bibitem[{{Aschwanden}(2004)}]{aschwandencorona}
{Aschwanden}, M.~J. 2004, {Physics of the Solar Corona. An Introduction}
  (Springer)

\bibitem[{{Astropy Collaboration} {et~al.}(2018){Astropy Collaboration},
  {Price-Whelan}, {Sip{\H{o}}cz}, {G{\"u}nther}, {Lim}, {Crawford}, {Conseil},
  {Shupe}, {Craig}, {Dencheva}, {Ginsburg}, {VanderPlas}, {Bradley},
  {P{\'e}rez-Su{\'a}rez}, {de Val-Borro}, {Aldcroft}, {Cruz}, {Robitaille},
  {Tollerud}, {Ardelean}, {Babej}, {Bach}, {Bachetti}, {Bakanov}, {Bamford},
  {Barentsen}, {Barmby}, {Baumbach}, {Berry}, {Biscani}, {Boquien}, {Bostroem},
  {Bouma}, {Brammer}, {Bray}, {Breytenbach}, {Buddelmeijer}, {Burke},
  {Calderone}, {Cano Rodr{\'\i}guez}, {Cara}, {Cardoso}, {Cheedella}, {Copin},
  {Corrales}, {Crichton}, {D'Avella}, {Deil}, {Depagne}, {Dietrich}, {Donath},
  {Droettboom}, {Earl}, {Erben}, {Fabbro}, {Ferreira}, {Finethy}, {Fox},
  {Garrison}, {Gibbons}, {Goldstein}, {Gommers}, {Greco}, {Greenfield},
  {Groener}, {Grollier}, {Hagen}, {Hirst}, {Homeier}, {Horton}, {Hosseinzadeh},
  {Hu}, {Hunkeler}, {Ivezi{\'c}}, {Jain}, {Jenness}, {Kanarek}, {Kendrew},
  {Kern}, {Kerzendorf}, {Khvalko}, {King}, {Kirkby}, {Kulkarni}, {Kumar},
  {Lee}, {Lenz}, {Littlefair}, {Ma}, {Macleod}, {Mastropietro}, {McCully},
  {Montagnac}, {Morris}, {Mueller}, {Mumford}, {Muna}, {Murphy}, {Nelson},
  {Nguyen}, {Ninan}, {N{\"o}the}, {Ogaz}, {Oh}, {Parejko}, {Parley}, {Pascual},
  {Patil}, {Patil}, {Plunkett}, {Prochaska}, {Rastogi}, {Reddy Janga},
  {Sabater}, {Sakurikar}, {Seifert}, {Sherbert}, {Sherwood-Taylor}, {Shih},
  {Sick}, {Silbiger}, {Singanamalla}, {Singer}, {Sladen}, {Sooley},
  {Sornarajah}, {Streicher}, {Teuben}, {Thomas}, {Tremblay}, {Turner},
  {Terr{\'o}n}, {van Kerkwijk}, {de la Vega}, {Watkins}, {Weaver}, {Whitmore},
  {Woillez}, {Zabalza}, \& {Astropy Contributors}}]{astropy:2018}
{Astropy Collaboration}, {Price-Whelan}, A.~M., {Sip{\H{o}}cz}, B.~M., {et~al.}
  2018, \aj, 156, 123

\bibitem[{{Astropy Collaboration} {et~al.}(2013){Astropy Collaboration},
  {Robitaille}, {Tollerud}, {Greenfield}, {Droettboom}, {Bray}, {Aldcroft},
  {Davis}, {Ginsburg}, {Price-Whelan}, {Kerzendorf}, {Conley}, {Crighton},
  {Barbary}, {Muna}, {Ferguson}, {Grollier}, {Parikh}, {Nair}, {Unther},
  {Deil}, {Woillez}, {Conseil}, {Kramer}, {Turner}, {Singer}, {Fox}, {Weaver},
  {Zabalza}, {Edwards}, {Azalee Bostroem}, {Burke}, {Casey}, {Crawford},
  {Dencheva}, {Ely}, {Jenness}, {Labrie}, {Lim}, {Pierfederici}, {Pontzen},
  {Ptak}, {Refsdal}, {Servillat}, \& {Streicher}}]{astropy:2013}
{Astropy Collaboration}, {Robitaille}, T.~P., {Tollerud}, E.~J., {et~al.} 2013,
  \aap, 558, A33

\bibitem[{{Barczynski} {et~al.}(2021){Barczynski}, {Schmieder}, {Peat},
  {Labrosse}, {Mein}, \& {Mein}}]{kb2021}
{Barczynski}, K., {Schmieder}, B., {Peat}, A.~W., {et~al.} 2021, arXiv
  e-prints, arXiv:2106.04259

\bibitem[{{De Pontieu} {et~al.}(2014){De Pontieu}, {Title}, {Lemen}, {Kushner},
  {Akin}, {Allard}, {Berger}, {Boerner}, {Cheung}, {Chou}, {Drake}, {Duncan},
  {Freeland}, {Heyman}, {Hoffman}, {Hurlburt}, {Lindgren}, {Mathur}, {Rehse},
  {Sabolish}, {Seguin}, {Schrijver}, {Tarbell}, {W{\"u}lser}, {Wolfson},
  {Yanari}, {Mudge}, {Nguyen-Phuc}, {Timmons}, {van Bezooijen}, {Weingrod},
  {Brookner}, {Butcher}, {Dougherty}, {Eder}, {Knagenhjelm}, {Larsen},
  {Mansir}, {Phan}, {Boyle}, {Cheimets}, {DeLuca}, {Golub}, {Gates}, {Hertz},
  {McKillop}, {Park}, {Perry}, {Podgorski}, {Reeves}, {Saar}, {Testa}, {Tian},
  {Weber}, {Dunn}, {Eccles}, {Jaeggli}, {Kankelborg}, {Mashburn}, {Pust},
  {Springer}, {Carvalho}, {Kleint}, {Marmie}, {Mazmanian}, {Pereira}, {Sawyer},
  {Strong}, {Worden}, {Carlsson}, {Hansteen}, {Leenaarts}, {Wiesmann},
  {Aloise}, {Chu}, {Bush}, {Scherrer}, {Brekke}, {Martinez-Sykora}, {Lites},
  {McIntosh}, {Uitenbroek}, {Okamoto}, {Gummin}, {Auker}, {Jerram}, {Pool}, \&
  {Waltham}}]{iris}
{De Pontieu}, B., {Title}, A.~M., {Lemen}, J.~R., {et~al.} 2014, \solphys, 289,
  2733

\bibitem[{{Driesman} {et~al.}(2008){Driesman}, {Hynes}, \& {Cancro}}]{stereo}
{Driesman}, A., {Hynes}, S., \& {Cancro}, G. 2008, \ssr, 136, 17

\bibitem[{{Golub} {et~al.}(2007){Golub}, {Deluca}, {Austin}, {Bookbinder},
  {Caldwell}, {Cheimets}, {Cirtain}, {Cosmo}, {Reid}, {Sette}, {Weber},
  {Sakao}, {Kano}, {Shibasaki}, {Hara}, {Tsuneta}, {Kumagai}, {Tamura},
  {Shimojo}, {McCracken}, {Carpenter}, {Haight}, {Siler}, {Wright}, {Tucker},
  {Rutledge}, {Barbera}, {Peres}, \& {Varisco}}]{xrt}
{Golub}, L., {Deluca}, E., {Austin}, G., {et~al.} 2007, \solphys, 243, 63

\bibitem[{{Gouttebroze} {et~al.}(1993){Gouttebroze}, {Heinzel}, \&
  {Vial}}]{GHV}
{Gouttebroze}, P., {Heinzel}, P., \& {Vial}, J.~C. 1993, \aaps, 99, 513

\bibitem[{{Gun{\'a}r} {et~al.}(2008){Gun{\'a}r}, {Heinzel}, {Anzer}, \&
  {Schmieder}}]{2008Gunar}
{Gun{\'a}r}, S., {Heinzel}, P., {Anzer}, U., \& {Schmieder}, B. 2008, \aap,
  490, 307

\bibitem[{{Gun{\'a}r} {et~al.}(2007){Gun{\'a}r}, {Heinzel}, {Schmieder},
  {Schwartz}, \& {Anzer}}]{2007Gunar}
{Gun{\'a}r}, S., {Heinzel}, P., {Schmieder}, B., {Schwartz}, P., \& {Anzer}, U.
  2007, \aap, 472, 929

\bibitem[{Harris {et~al.}(2020)Harris, Millman, van~der Walt, Gommers,
  Virtanen, Cournapeau, Wieser, Taylor, Berg, Smith, Kern, Picus, Hoyer, van
  Kerkwijk, Brett, Haldane, Fernández~del Río, Wiebe, Peterson,
  Gérard-Marchant, Sheppard, Reddy, Weckesser, Abbasi, Gohlke, \&
  Oliphant}]{numpy}
Harris, C.~R., Millman, K.~J., van~der Walt, S.~J., {et~al.} 2020, Nature, 585,
  357–362

\bibitem[{{Heinzel} {et~al.}(1994){Heinzel}, {Gouttebroze}, \& {Vial}}]{HGV}
{Heinzel}, P., {Gouttebroze}, P., \& {Vial}, J.~C. 1994, \aap, 292, 656

\bibitem[{{Heinzel} {et~al.}(2015){Heinzel}, {Schmieder}, {Mein}, \&
  {Gun{\'a}r}}]{heinzel2015}
{Heinzel}, P., {Schmieder}, B., {Mein}, N., \& {Gun{\'a}r}, S. 2015, \apjl,
  800, L13

\bibitem[{{Heinzel} {et~al.}(2014){Heinzel}, {Vial}, \& {Anzer}}]{heinzel2014}
{Heinzel}, P., {Vial}, J.~C., \& {Anzer}, U. 2014, \aap, 564, A132

\bibitem[{{Howard} {et~al.}(2008){Howard}, {Moses}, {Vourlidas}, {Newmark},
  {Socker}, {Plunkett}, {Korendyke}, {Cook}, {Hurley}, {Davila}, {Thompson},
  {St Cyr}, {Mentzell}, {Mehalick}, {Lemen}, {Wuelser}, {Duncan}, {Tarbell},
  {Wolfson}, {Moore}, {Harrison}, {Waltham}, {Lang}, {Davis}, {Eyles},
  {Mapson-Menard}, {Simnett}, {Halain}, {Defise}, {Mazy}, {Rochus}, {Mercier},
  {Ravet}, {Delmotte}, {Auchere}, {Delaboudiniere}, {Bothmer}, {Deutsch},
  {Wang}, {Rich}, {Cooper}, {Stephens}, {Maahs}, {Baugh}, {McMullin}, \&
  {Carter}}]{secchi}
{Howard}, R.~A., {Moses}, J.~D., {Vourlidas}, A., {et~al.} 2008, \ssr, 136, 67

\bibitem[{Hunter(2007)}]{matplotlib}
Hunter, J.~D. 2007, Computing in Science \& Engineering, 9, 90

\bibitem[{{Hyder} \& {Lites}(1970)}]{1970hyder}
{Hyder}, C.~L. \& {Lites}, B.~W. 1970, \solphys, 14, 147

\bibitem[{{Jej{\v{c}}i{\v{c}}} {et~al.}(2018){Jej{\v{c}}i{\v{c}}}, {Schwartz},
  {Heinzel}, {Zapi{\'o}r}, \& {Gun{\'a}r}}]{jejcic2018}
{Jej{\v{c}}i{\v{c}}}, S., {Schwartz}, P., {Heinzel}, P., {Zapi{\'o}r}, M., \&
  {Gun{\'a}r}, S. 2018, \aap, 618, A88

\bibitem[{{Kerr} {et~al.}(2015){Kerr}, {Sim{\~o}es}, {Qiu}, \&
  {Fletcher}}]{kerr2015}
{Kerr}, G.~S., {Sim{\~o}es}, P.~J.~A., {Qiu}, J., \& {Fletcher}, L. 2015, \aap,
  582, A50

\bibitem[{{Kosugi} {et~al.}(2007){Kosugi}, {Matsuzaki}, {Sakao}, {Shimizu},
  {Sone}, {Tachikawa}, {Hashimoto}, {Minesugi}, {Ohnishi}, {Yamada}, {Tsuneta},
  {Hara}, {Ichimoto}, {Suematsu}, {Shimojo}, {Watanabe}, {Shimada}, {Davis},
  {Hill}, {Owens}, {Title}, {Culhane}, {Harra}, {Doschek}, \&
  {Golub}}]{hinodeoverview}
{Kosugi}, T., {Matsuzaki}, K., {Sakao}, T., {et~al.} 2007, \solphys, 243, 3

\bibitem[{{Kucera} {et~al.}(2018){Kucera}, {Ofman}, \& {Tarbell}}]{kucera2018}
{Kucera}, T.~A., {Ofman}, L., \& {Tarbell}, T.~D. 2018, \apj, 859, 121

\bibitem[{{Labrosse} \& {Gouttebroze}(2004)}]{promctr}
{Labrosse}, N. \& {Gouttebroze}, P. 2004, \apj, 617, 614

\bibitem[{{Labrosse} {et~al.}(2007){Labrosse}, {Gouttebroze}, \&
  {Vial}}]{2007labrosse}
{Labrosse}, N., {Gouttebroze}, P., \& {Vial}, J.~C. 2007, \aap, 463, 1171

\bibitem[{{Labrosse} {et~al.}(2010){Labrosse}, {Heinzel}, {Vial}, {Kucera},
  {Parenti}, {Gun{\'a}r}, {Schmieder}, \& {Kilper}}]{promreviewi}
{Labrosse}, N., {Heinzel}, P., {Vial}, J.~C., {et~al.} 2010, \ssr, 151, 243

\bibitem[{{Labrosse} \& {Rodger}(2016)}]{2016Rodger}
{Labrosse}, N. \& {Rodger}, A.~S. 2016, \aap, 587, A113

\bibitem[{{Leenaarts} {et~al.}(2013){Leenaarts}, {Pereira}, {Carlsson},
  {Uitenbroek}, \& {De Pontieu}}]{irisii}
{Leenaarts}, J., {Pereira}, T.~M.~D., {Carlsson}, M., {Uitenbroek}, H., \& {De
  Pontieu}, B. 2013, \apj, 772, 90

\bibitem[{{Lemen} {et~al.}(2012){Lemen}, {Title}, {Akin}, {Boerner}, {Chou},
  {Drake}, {Duncan}, {Edwards}, {Friedlaender}, {Heyman}, {Hurlburt}, {Katz},
  {Kushner}, {Levay}, {Lindgren}, {Mathur}, {McFeaters}, {Mitchell}, {Rehse},
  {Schrijver}, {Springer}, {Stern}, {Tarbell}, {Wuelser}, {Wolfson}, {Yanari},
  {Bookbinder}, {Cheimets}, {Caldwell}, {Deluca}, {Gates}, {Golub}, {Park},
  {Podgorski}, {Bush}, {Scherrer}, {Gummin}, {Smith}, {Auker}, {Jerram},
  {Pool}, {Soufli}, {Windt}, {Beardsley}, {Clapp}, {Lang}, \& {Waltham}}]{aia}
{Lemen}, J.~R., {Title}, A.~M., {Akin}, D.~J., {et~al.} 2012, \solphys, 275, 17

\bibitem[{{Levens} \& {Labrosse}(2019)}]{levensmgii}
{Levens}, P.~J. \& {Labrosse}, N. 2019, \aap, 625, A30

\bibitem[{{Levens} {et~al.}(2016){Levens}, {Schmieder}, {Labrosse}, \&
  {L{\'o}pez Ariste}}]{levenspromlegs}
{Levens}, P.~J., {Schmieder}, B., {Labrosse}, N., \& {L{\'o}pez Ariste}, A.
  2016, \apj, 818, 31

\bibitem[{{Liggett} \& {Zirin}(1984)}]{liggett1984}
{Liggett}, M. \& {Zirin}, H. 1984, \solphys, 91, 259

\bibitem[{{Liu} {et~al.}(2015){Liu}, {De Pontieu}, {Vial}, {Title}, {Carlsson},
  {Uitenbroek}, {Okamoto}, {Berger}, \& {Antolin}}]{liu2015}
{Liu}, W., {De Pontieu}, B., {Vial}, J.-C., {et~al.} 2015, \apj, 803, 85

\bibitem[{{Mackay} {et~al.}(2010){Mackay}, {Karpen}, {Ballester}, {Schmieder},
  \& {Aulanier}}]{promreviewii}
{Mackay}, D.~H., {Karpen}, J.~T., {Ballester}, J.~L., {Schmieder}, B., \&
  {Aulanier}, G. 2010, \ssr, 151, 333

\bibitem[{{Mein}(1991)}]{mein1991}
{Mein}, P. 1991, \aap, 248, 669

\bibitem[{{Parenti} {et~al.}(2012){Parenti}, {Schmieder}, {Heinzel}, \&
  {Golub}}]{parenti2012}
{Parenti}, S., {Schmieder}, B., {Heinzel}, P., \& {Golub}, L. 2012, \apj, 754,
  66

\bibitem[{Pereira {et~al.}(2018)Pereira, McIntosh, De~Pontieu, Hansteen,
  Carlsson, Boerner, Liu, Gošic', \& Green}]{itn26}
Pereira, T.~M., McIntosh, S.~W., De~Pontieu, B., {et~al.} 2018, ITN 26: A users
  guide to IRIS data retrieval reduction \& analysis,
  \url{https://iris.lmsal.com/itn26/}

\bibitem[{{Pesnell} {et~al.}(2012){Pesnell}, {Thompson}, \& {Chamberlin}}]{sdo}
{Pesnell}, W.~D., {Thompson}, B.~J., \& {Chamberlin}, P.~C. 2012, \solphys,
  275, 3

\bibitem[{{Ruan} {et~al.}(2019){Ruan}, {Jej{\v{c}}i{\v{c}}}, {Schmieder},
  {Mein}, {Mein}, {Heinzel}, {Gun{\'a}r}, \& {Chen}}]{ruan2019}
{Ruan}, G., {Jej{\v{c}}i{\v{c}}}, S., {Schmieder}, B., {et~al.} 2019, \apj,
  886, 134

\bibitem[{{Ruan} {et~al.}(2018){Ruan}, {Schmieder}, {Mein}, {Mein}, {Labrosse},
  {Gun{\'a}r}, \& {Chen}}]{ruan2018}
{Ruan}, G., {Schmieder}, B., {Mein}, P., {et~al.} 2018, \apj, 865, 123

\bibitem[{{Schmieder} {et~al.}(2017){Schmieder}, {Zapi{\'o}r}, {L{\'o}pez
  Ariste}, {Levens}, {Labrosse}, \& {Gravet}}]{levens3d}
{Schmieder}, B., {Zapi{\'o}r}, M., {L{\'o}pez Ariste}, A., {et~al.} 2017, \aap,
  606, A30

\bibitem[{{Tandberg-Hanssen}(1995)}]{hanssen1995}
{Tandberg-Hanssen}, E. 1995, {The nature of solar prominences}, Vol. 199
  (Kluwer Academic Publishers)

\bibitem[{{The SunPy Community} {et~al.}(2020){The SunPy Community}, Barnes,
  Bobra, Christe, Freij, Hayes, Ireland, Mumford, Perez-Suarez, Ryan, Shih,
  Chanda, Glogowski, Hewett, Hughitt, Hill, Hiware, Inglis, Kirk, Konge, Mason,
  Maloney, Murray, Panda, Park, Pereira, Reardon, Savage, Sip{\H{o}}cz,
  Stansby, Jain, Taylor, Yadav, Rajul, Dang, \& and}]{sunwhy}
{The SunPy Community}, Barnes, W.~T., Bobra, M.~G., {et~al.} 2020, The
  Astrophysical Journal, 890, 68

\bibitem[{{Vial}(1998)}]{vial1998}
{Vial}, J.~C. 1998, in Astronomical Society of the Pacific Conference Series,
  Vol. 150, IAU Colloq. 167: New Perspectives on Solar Prominences, ed. D.~F.
  {Webb}, B.~{Schmieder}, \& D.~M. {Rust}, 175

\bibitem[{Vial \& Engvold(2015)}]{newprom}
Vial, J.-C. \& Engvold, O. 2015, Solar Prominences, Vol. 415;415.; (Cham:
  Springer International Publishing)

\bibitem[{Virtanen {et~al.}(2020)Virtanen, Gommers, Oliphant, Haberland, Reddy,
  Cournapeau, Burovski, Peterson, Weckesser, Bright, {van der Walt}, Brett,
  Wilson, Millman, Mayorov, Nelson, Jones, Kern, Larson, Carey, Polat, Feng,
  Moore, {VanderPlas}, Laxalde, Perktold, Cimrman, Henriksen, Quintero, Harris,
  Archibald, Ribeiro, Pedregosa, {van Mulbregt}, \& {SciPy 1.0
  Contributors}}]{scipy}
Virtanen, P., Gommers, R., Oliphant, T.~E., {et~al.} 2020, Nature Methods, 17,
  261

\bibitem[{{W{\"u}lser} {et~al.}(2018){W{\"u}lser}, {Jaeggli}, {De Pontieu},
  {Tarbell}, {Boerner}, {Freeland}, {Liu}, {Timmons}, {Brannon}, {Kankelborg},
  {Madsen}, {McKillop}, {Prchlik}, {Saar}, {Schanche}, {Testa}, {Bryans}, \&
  {Wiesmann}}]{iriscal2018}
{W{\"u}lser}, J.~P., {Jaeggli}, S., {De Pontieu}, B., {et~al.} 2018, \solphys,
  293, 149

\bibitem[{{Zhang} {et~al.}(2019){Zhang}, {Buchlin}, \&
  {Vial}}]{zhangcmepromiris}
{Zhang}, P., {Buchlin}, {\'E}., \& {Vial}, J.~C. 2019, \aap, 624, A72

\end{thebibliography}
\bibliographystyle{aa}
\onecolumn 
\appendix
\section{\ion{Mg}{ii} k diagnostic maps}

\begin{figure}[ht]
    \centering
    \includegraphics[width=\textwidth]{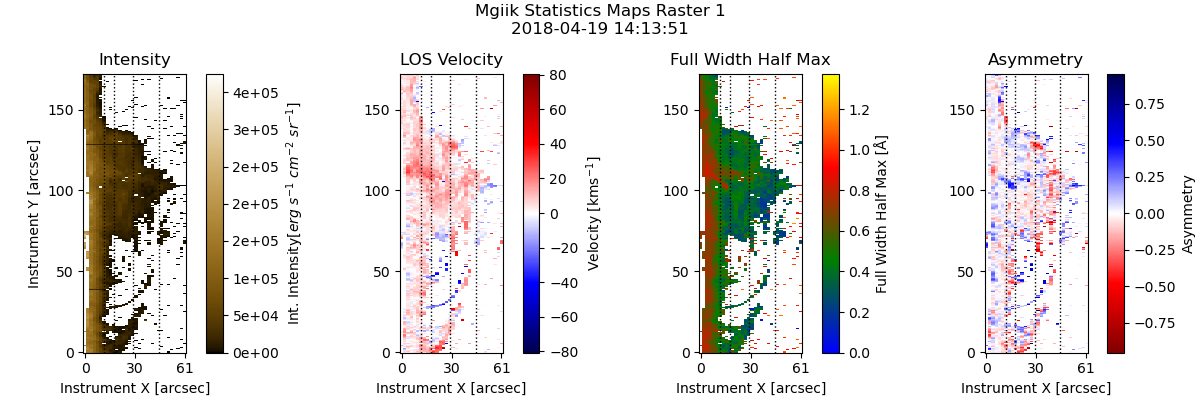}\\
    \includegraphics[width=\textwidth]{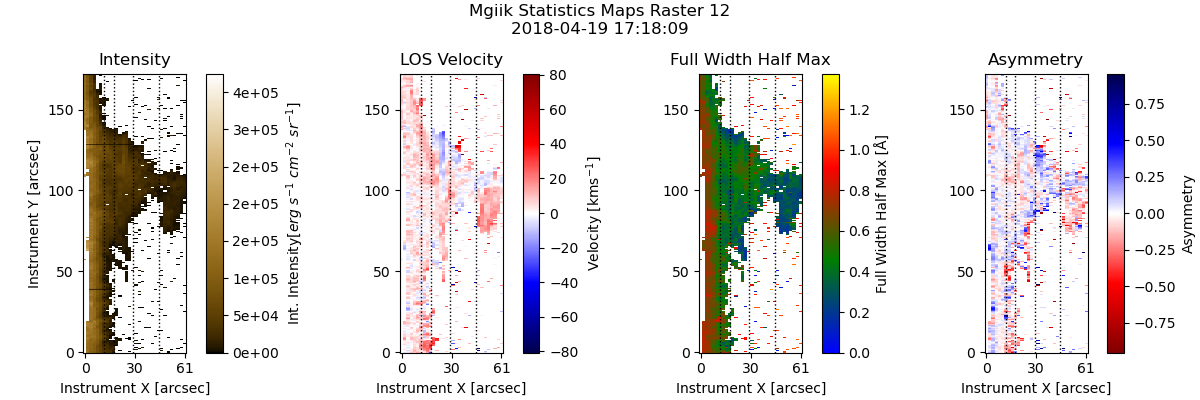}\\
    \includegraphics[width=\textwidth]{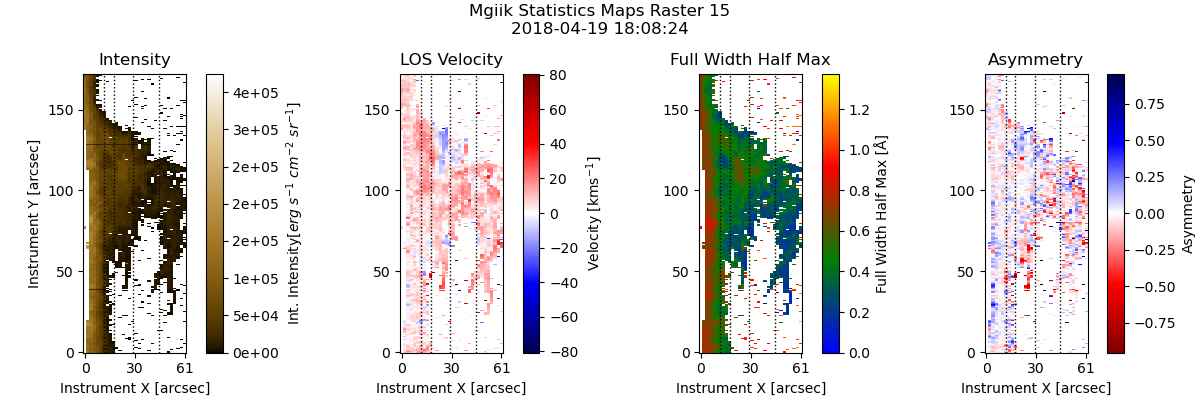}
    \caption[width=\textwidth]{\ion{Mg}{ii} k statistics maps of the three main stages of the prominence observation calculated via the quantile method. The times associated with these plots are the time at the beginning of the associated raster. The dashed lines represent the slit positions in Figure~\ref{slitevo}. An animated version of this figure is available online.}
    \label{kmaps}
\end{figure}

\end{document}